\documentclass{emulateapj}

\def\simgt{\lower 2pt \hbox{$\, \buildrel {\scriptstyle >}\over{\scriptstyle \sim}\,$}}
\def\simlt{\lower 2pt \hbox{$\, \buildrel {\scriptstyle <}\over{\scriptstyle \sim}\,$}}

\begin{document}

\title{Modeling the Accretion Structure of AU Mon}

\author{Corwin Atwood-Stone\altaffilmark{1,2}, Brendan P. Miller\altaffilmark{3,1}, Mercedes T. Richards\altaffilmark{4}, J{\'a}n Budaj\altaffilmark{5}, and Geraldine J. Peters\altaffilmark{6}}

\altaffiltext{1}{Department of Physics, College of Wooster, Wooster, OH 44691, USA}
\altaffiltext{2}{Lunar and Planetary Laboratory, University of Arizona, Tucson, AZ 85721, USA; catwoods@lpl.arizona.edu}
\altaffiltext{3}{Department of Astronomy, University of Michigan, Ann Arbor, MI 48109, USA; mbrendan@umich.edu}
\altaffiltext{4}{Department of Astronomy \& Astrophysics, Pennsylvania State University, University Park, PA 16802, USA; mrichards@astro.psu.edu}
\altaffiltext{5}{Astronomical Institute, Slovak Academy of Sciences, Tatranska Lomnica 05960, Slovak Republic; budaj@ta3.sk}
\altaffiltext{6}{Space Sciences Center and Department of Physics \& Astronomy, 
University of Southern California, Los Angeles, CA 90089-1341, USA; gjpeters@mucen.usc.edu}

\begin{abstract}

AU Mon is a long-period (11.113 d) Algol-type binary system with a
persistent accretion disk that is apparent as double-peaked H$\alpha$
emission. We present previously unpublished optical spectra of AU Mon
which were obtained over several years with dense orbital phase
coverage.  We utilize these data, along with archival UV spectra, to
model the temperature and structure of the accretion disk and the gas
stream.  Synthetic spectral profiles for lines including H$\alpha$,
H$\beta$, and the \ion{Al}{3} and \ion{Si}{4} doublets were computed
with the {\it Shellspec\/} program.  The best match between the model 
spectra and the observations is obtained for an accretion disk of inner/outer radius 5.1/23
$R_{\odot}$, thickness of 5.2$R_{\odot}$, density of $1.0\times10^{-13}$~g~cm$^{-3}$, 
and maximum temperature of 14000 K, along with a gas stream at a temperature of
$\sim$8000 K transferring $\sim$$2.4\times10^{-9} M_{\odot}$yr$^{-1}$.  
We show H$\alpha$ Doppler tomograms of the velocity structure of the gas,
constructed from difference profiles calculated through sequentially
subtracting contributions from the stars and accretion structures. The
tomograms provide independent support for the {\it Shellspec\/}
modeling, while also illustrating that residual emission at
sub-Keplerian velocities persists even after subtracting the disk and
stream emission. Spectral variability in the H$\alpha$ profile beyond
that expected from either the orbital or the long-period cycle is
present on both multi-week and multi-year timescales, and may reflect
quasi-random changes in the mass transfer rate or the disk
structure. Finally, a transient UV spectral absorption feature may be
modeled as an occasional outflow launched from the vicinity of the
disk-stream interaction region.

\end{abstract}



\section{Introduction}

Algol-type eclipsing binary systems consist of semi-detached
components in which the secondary has evolved to overflow its Roche
lobe.  The current mass ratio results from the transfer of sufficient material 
onto the less-evolved primary until it was transformed into the more massive 
of the two stars (Crawford 1955; Kopal 1955). The
relative separation of the components, the radius of the
primary, and the mass ratio determine the types of accretion structures 
that eventually form. In short-period Algols ($P\simlt4.5$ days), the gas stream
arising from the L1 point can impact the gainer directly; in
long-period systems ($P\simgt6$ days), the curving gas stream may
instead feed into a stable disk; and in intermediate-period systems
mixed behavior has been observed (e.g., Peters 1989, Richards \&
Albright 1999, and references therein). Secondary structures such as
hotspots, outflows (potentially including jets), or accretion annuli
can also develop. AU~Mon (GCRV~4526, HD~50846, HIP~33237), with an orbital period of 11.113 d, has one of the longer periods among the well-studied Algol-type binaries 
(Richards \& Albright 1999; Desmet et al.~2010), and is consequently
ideally suited for investigation of the properties of an enduring accretion disk (even though it is sometimes variable), along with those of the gas stream and other accretion structures.

The components of Algol-type binary systems are typically too close to
resolve as individual stars.  A notable exception is the bright and
nearby system of Algol ($\beta$ Per) for which optical interferometry
was used recently to resolve all three stars in that system (Zavala et
al.~2010), with orbital parameters confirmed and refined by radio
interferometry (Peterson et al.~2011).  These direct imaging techniques
are still not able to resolve any of the non-stellar components
including the accretion disk and gas stream.  Instead, the non-stellar
components can be revealed more readily with spectroscopy by
disentangling the contributions of the stars from the composite
spectrum. The receding/approaching sides of an accretion disk manifest
as red/blue-shifted emission superposed on the stellar absorption line
profile, as first noted by Joy (1942). The persistent presence of
double-peaked H$\alpha$ emission in the composite spectrum serves as
one of the more prominent observational indicators of an enduring
accretion disk. Additional accretion features, such as the gas stream,
also influence the spectral profile to a lesser extent (e.g., as
excess redshifted absorption when the viewpoint is through the
stream, and as emission when viewed across the stream). A
two-dimensional velocity mapping of the gas in the stream and disk may
be constructed through Doppler tomography of the line profiles
obtained at orbital phases around the binary (e.g., Marsh \& Horne~1988; review by
Richards 2004, and references therein). This technique has been
extended to provide three-dimensional velocity maps (e.g., Agafonov et
al.~2009; Richards et al.~2010; Richards et al.~2012). Hydrodynamic simulations provide
another method of investigating the distribution of accreting gas, and
can reproduce many of the features seen in the tomograms (Richards \&
Ratliff 1998; Bisikalo \& Kononov 2010; Raymer 2012).

A complementary approach involves computing the spectral contribution
from geometric representations of the accretion structures, so that
the physical characteristics of the disk, stream, or other features
(size scales, density, temperature) may be determined through matching
the modeled synthetic spectra to the observations. In this work, we make
use of the {\it Shellspec\/}
%{\footnote{http://www.astro.sk/$\sim$budaj/shellspec.html}}
program developed by Budaj \& Richards (2004). Briefly, this program
calculates the line-of-sight radiative transfer within the moving
circumstellar environment, and the stars can have Roche model geometries.  
For a given set of phases, a composite spectrum of the
interacting binary is calculated with associated accretion structures:
optionally including an accretion disk, a gas stream, a spot, a jet,
and a shell.  Consideration of multiple transitions helps to break
potential parameter space degeneracies in modeling. Since Algol-type
binaries frequently display variable line profiles, plausibly due to
changes in the rate of mass transfer (e.g., Plavec \& Polidan 1976;
Richards \& Albright 1999; Richards 2004), the parameters determined
from the comparison of models with the data at a particular epoch
describe the then-dominant accretion mode, and multi-epoch
observations are required to explore the full range of conditions
occurring within the system.

{\it Shellspec\/} provides a flexible framework for modification and
extension to related astrophysical topics; for example, Tkachenko et
al.~(2009, 2010) created an inverted version of the code to determine properties of
the component stars; Budaj (2011a) described an improved consideration
of the reflection effect in {\it Shellspec\/} which he applied to
modeling the temperature distribution over the surface of transiting
exoplanets; Budaj (2011b) used {\it Shellspec\/} to model dust disks in
$\epsilon$~Aur; and Chadima et al.~(2011) used {\it Shellspec\/} along
with ZEUS-MP hydrodynamic models to illustrate the phase-varying
H$\alpha$ V/R asymmetry resulting from injection of a blob of gaseous
material (such as could occur from discontinuous mass transfer prior
to viscosity-driven smoothing).

The current version of {\it Shellspec\/} applies a straight cylindrical
representation of the gas stream and a circularly symmetric accretion
disk, and the user may upload a general 3D accretion structure with
optional velocity, temperature, and density fields (see {www.astro.sk/$\sim$budaj/shellspec.html}). Accretion disks in
Algol-type binaries have often been suggested to lack circular symmetry or to possess regions of enhanced emission beyond that expected from a symmetric disk (e.g., the asymmetric disk model applied to RY Per by Barai et al.~(2004), although their method did not include radiative transfer).
{\it Shellspec\/} modeling of the accretion structures in TT Hya by Budaj et al.~(2005) and Miller et al.~(2007) demonstrated that the observed spectra could not be reproduced solely by the gas stream and a symmetric disk.  In these cases, {\it Shellspec\/} can reproduce the observed characteristics through inclusion of the ``spot'' feature.  

In this work, we investigate the partially-eclipsing long-period Algol-type
binary AU Mon. Comprehensive studies of the components of AU Mon based
on CoRoT photometry were recently performed by Desmet et al.~(2010)
and Djura{\v s}evi{\'c} et al.~(2010), with the former work including
high-resolution spectroscopy to help constrain the stellar properties
and the latter work additionally modeling the accretion disk; we use
their results for system elements. Prior determinations of system
parameters were made by Sahade et al.~(1997) and Vivekananda
Rao \& Sarma~(1998), and values for the ephemeris and period of
AU Mon were derived by Lorenzi (1980) and Kreiner (2004). As in
many Algol-type binary systems, the primary is carbon-poor, presumably
due to deposition of CNO-processed material (Ibanoglu et
al.~2012). Double-peaked H$\alpha$ emission has been seen in all
observations (e.g., Plavec \& Polidan 1976; Peters 1989; Sahade et
al.~1996, 1997; Richards \& Albright 1999; Desmet et al.~2010;
Barr{\'{\i}}a \& Mennickent~2011) with a profile that may vary between
epochs (e.g., Plavec \& Polidan 1976; Sahade et al.~1997). Peters \&
Polidan (1998) inferred a particle density in the disk of
$10^{8-9}$~cm$^{-3}$ based on an examination of FUV \ion{Fe}{3}
features. Peters \& Polidan (1984, 1998) found evidence from UV
spectra of a high-temperature accretion region (HTAR) in several
Algol-type binaries, including AU Mon. They determined that the HTAR
was located near the trailing hemisphere of the primary (i.e., in the
vicinity of the impact region where the gas stream collides with the 
stellar surface), with an electron
temperature and density of around $T_{\rm e}=10^{5}$~K and $n_{\rm
e}=10^{9}$~cm$^{-3}$, and was carbon-depleted (which suggested
prior CNO processing of this material). Most recently,
Djura{\v s}evi{\'c} et al.~(2010) re-analyzed CoRoT and $V$-band
photometry using a Roche model with a non-transparent accretion
disk. They found that a ``hot spot'' and two ``bright spots''
(associated with spiral arms or deviations from a circular shape)
account for the observed asymmetry in the light curve of AU Mon. We
discuss their model in more detail in $\S$3.1. Mimica \&
Pavlovski~(2012), also used CoRoT photometry to independently
characterize the disk of AU~Mon as possessing a ``clumpy'' structure.

AU Mon was discovered by Lorenzi (1980) to display a long-period variation in brightness of $\sim$0.2 magnitudes, with a period of $\sim$417 days (or 37.5 times the orbital period).  This characteristic has now been associated with the Double Periodic Variables (DPV; Mennickent et al. 2003) that show two photometric periods, one associated with orbital motion and the other a cyclic long-term light modulation that is about 35 times the orbital period. These systems tend to be more luminous than classical Algols and are categorized by mass loss and circumbinary disks in addition to mass transfer between stars; hence they can be used to provide constraints on non-conservative models of binary star evolution (Mennickent et al. 2008).  AU Mon was one of the first DPV systems discovered in our Galaxy within the class of Algol-type binaries; its properties will be compared with V393 Sco, a recently discovered DPV system (Mennickent et al.~2010).

Desmet et al.~(2010) found that the shape of the light curves of AU Mon during long-period maxima and minima are identical but offset, and suggested that the intensity variations are instead linked to changes in the attenuation from circumbinary material; this interpretation is in agreement with Djura{\v s}evi{\'c} et al.~(2010).   Mennickent et al.~(2008) hypothesized that
variable mass outflow alters the amount of circumbinary material in DPVs, and
Mennickent et al.~(2010) associated the long-period variability in V393 Sco
with equatorial mass loss. However, Peters (1994) found
that the UV absorption associated with the gas stream was stronger
when the stars were faint, and ascribed the long-period variation to
changes in the mass transfer rate due to cyclic expansion and
contraction of the secondary. Consistent with this, Barr{\'{\i}}a \&
Mennickent~(2011) showed that the H$\alpha$ and H$\beta$ central
absorption in AU Mon is likewise deeper near and following minimum light, and
concluded that the profile variability may result from a changing
circumprimary envelope influenced by mass loss through the L1 point. 

This paper is organized as follows: $\S$2 describes the data and
observations, including previously unpublished high-resolution optical
spectra of AU Mon; $\S$3 contains the {\it Shellspec\/} modeling
process and the calculated characteristics of the accretion features;
$\S$4 considers spectral variability in AU Mon on both multi-year and 
multi-week timescales; $\S$5 presents the Doppler tomography results 
based on both observed and difference spectra, including comparisons 
between the observed spectra and the {\it Shellspec\/} model of the 
stars, disk, gas stream, and a high density localized region; 
$\S$6 provides a summary of our results.

%%%%%%%%%%%%%%%%%%%%%%%%%%%%%%%%%%%%%%%%%%%%%%%%%%%%%%%%%%%%%

%% Figure 1 HERE

\begin{figure}
\figurenum{1}
\epsscale{1.1}
\plotone{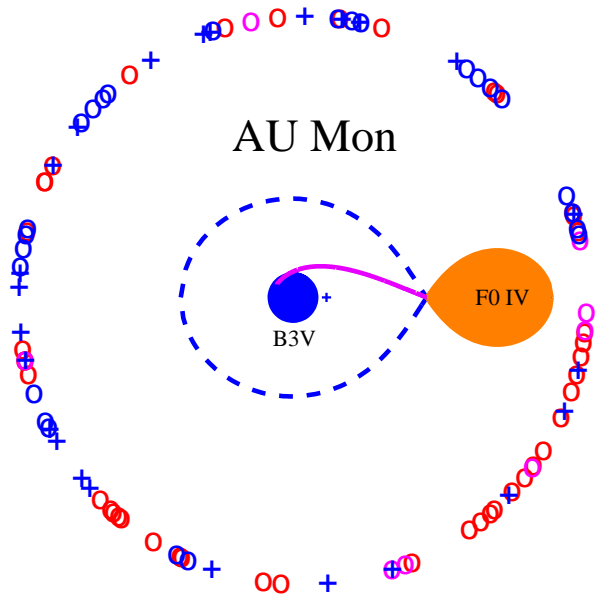}
\caption{Geometry and phase coverage of AU Mon (see also Table~1). The
dotted blue line is the primary's Roche lobe, the purple curve
illustrates the path of the gas stream, and the central plus sign
shows the location of the center of mass of the binary. The outer blue plus
signs mark the Richards HET data, while the circles are KPNO data:
blue is from Richards and red/purple are the first/second Peters
runs. The H$\alpha$ spectra provide dense coverage of the entire orbital cycle.}
\label{f1}
\end{figure}

\section{Observations}

The ninety-five (95) high-resolution optical spectra of AU Mon which
form the primary focus of this work were collected at four
observatories over a period of 20 years from 1991 to 2011: Kitt Peak
National Observatory (KPNO), Hobby Eberly Telescope (HET), MDM
Observatory at Kitt Peak, and Las Campanas Observatory (LCO).  (The MDM Observatory is operated by a consortium of five universities: the University of Michigan, Dartmouth College, the Ohio State University, Columbia University, and Ohio University.)  These observations
are presented here for the first time, with the exception of 13 KPNO
spectra. The observations are listed in Table 1.

Seventy-two (72) H$\alpha$ spectra of AU Mon were obtained at KPNO
during several observing runs from 1991 to 2001.  Most of these
spectra were obtained with the 0.9m Coud{\'e} Feed Telescope using grating
A, providing a resolution of $R\simeq$20,000 across a spectral field
of $\sim$345\AA. Thirteen (13) of these spectra were obtained in 1994
by Richards and were previously described in Richards \& Albright
(1999).  In addition, twenty-four (24) HET spectra were obtained with
the HET High-Resolution Spectrograph (HRS) over 42 nights in 2002
December and 2003 January, with the 316g5936 cross disperser at a
resolution of $R={\lambda}/{\Delta}{\lambda}\simeq$30,000 and cover
4200--8000~\AA.  The HET spectra include both the H$\alpha$ and the
H$\beta$ regions. Finally, six (6) current-epoch spectra were
obtained at MDM observatory (5 H$\alpha$ spectra, 2011 February) and
at Las Campanas (1 optical spectrum, 2011 November).  The MDM H$\alpha$
spectra were obtained with the 2.4m Hiltner telescope using the Boller
and Chivens CCD Spectrograph (CCDS) with the 1800 grating, providing a
resolution of R$\simeq$7,500 across a spectral field of $\sim$330\AA,
and the LCO optical spectra were obtained with the 6.5m Clay Magellan
telescope using the Magellan Inamori Kyocera {\'E}chelle (MIKE)
double-{\'E}chelle spectrograph, providing a resolution of R$\simeq$30,000
near H$\alpha$.

%%TABLE 1 HERE

\begin{deluxetable}{lrrrrrr}
\tabletypesize{\scriptsize}
\tablecaption{Optical Observations}
\tablewidth{9.2cm}
\tablehead{\colhead{ID\tablenotemark{a}} &\colhead{HJD} & \colhead{${\phi}_{\rm L}$} & 
\colhead{${\phi}_{\rm 0}$} & \colhead{$I_{\rm abs}$} & \colhead{$I_{\rm blue}$} & \colhead{$I_{\rm red}$} }
\startdata
%KPNO--Peters
KPNO1 & 2446122.728&	 7.24&	298.870&    	0.96&	1.17&	1.17 \\
 & 	2446734.952&	 8.71&	353.961&    	0.59&	1.13&	1.22 \\ 
 & 	2446734.862&	 8.71&	353.952&    	0.59&	1.09&	1.21 \\	
 & 	2446902.661&	 9.11&	369.052&    	0.72&	1.27&	1.16 \\
 & 	2446903.663&	 9.11&	369.142&    	1.05&	1.26& 	1.22 \\	 
 & 	2446905.643&	 9.12&	369.320&    	1.09&	1.27&	1.27 \\		
 & 	2446919.634&	 9.15&	370.579&    	1.00&	1.25&	1.28 \\
 & 	2447033.970&	 9.42&	380.868&    	0.76&	1.17&	1.24 \\
 & 	2447469.840&	10.47&	420.089&    	0.48&	1.19&	1.10 \\
 & 	2447470.931&	10.47&	420.187&    	0.65&	1.14&	1.10 \\
 & 	2447471.868&	10.47&	420.272&    	0.70&	1.13&	1.10 \\
 & 	2447472.878&	10.48&	420.362&    	0.68&	1.10&	1.16 \\
 & 	2447560.669&	10.69&	428.262&    	0.78&	1.28&	1.19 \\
 & 	2447561.755&	10.69&	428.360&    	0.81&	1.21&	1.25 \\   	
 & 	2447636.645&	10.87&	435.099&    	0.94&	1.31&	1.15 \\   	
 & 	2447639.632&	10.88&	435.368&    	1.08&	1.26&	1.26 \\
 & 	2447640.637&	10.88&	435.458&    	1.04&	1.25&	1.24 \\
 & 	2447939.662&	11.60&	462.366&    	0.70&	1.08&	1.14 \\
 & 	2447939.778&	11.60&	462.376&    	0.66&	1.11&	1.19 \\
 & 	2447940.780&	11.60&	462.466&    	0.66&	1.14&	1.17 \\
 & 	2447941.593&	11.60&	462.539&    	0.70&	1.08&	1.18 \\
 & 	2447981.716&	11.70&	466.150&    	0.81&	1.08&	1.08 \\
 & 	2447983.615&	11.70&	466.321&    	0.85&	1.11&	1.11 \\   	
 & 	2448121.971&	12.03&	478.771&    	1.04&	1.37&	1.23 \\
 & 	2448123.959&	12.04&	478.950&    	0.82&	1.20&	1.28 \\
 & 	2448124.982&	12.04&	479.042&    	0.71&	1.38&	1.19 \\
 & 	2448125.985&	12.04&	479.132&    	1.05&	1.35&	1.23 \\
 & 	2448313.706&	12.49&	496.024&    	0.34&	1.41&	1.17 \\
 & 	2448313.792&	12.49&	496.032&    	0.35&	1.34&	1.07 \\
 & 	2448313.642&	12.49&	496.018&    	0.33&	1.45&	1.21 \\
 & 	2448314.754&	12.50&	496.118&    	0.59&	1.26&	1.18 \\
 & 	2448314.634&	12.50&	496.107&    	0.56&	1.25&	1.16 \\
 & 	2448318.688&	12.51&	496.472&    	0.53&	1.24&	1.25 \\
 & 	2448318.621&	12.51&	496.466&    	0.56&	1.27&	1.24 \\
 & 	2448319.764&	12.51&	496.569&    	0.63&	1.23&	1.30 \\
 & 	2448319.764&	12.51&	496.569&    	0.63&	1.23&	1.30 \\
 & 	2448320.619&	12.51&	496.646&    	0.67&	1.27&	1.19 \\   
 & 	2448321.616&	12.51&	496.736&    	0.69&	1.17&	1.21 \\   	
 & 	2448514.977&	12.98&	514.135&    	1.01&	1.30&	1.19 \\   
 & 	2449058.775&	14.28&	563.068&    	0.64&	1.34&	1.24 \\   	
 & 	2449061.776&	14.29&	563.338&    	0.84&	1.31&	1.29 \\		
 & 	2449443.708&	15.20&	597.706&    	0.99&	1.21&	1.22 \\   	
 & 	2449444.690&	15.20&	597.795&    	1.03&	1.19&	1.14 \\
 & 	2449688.900&    15.79 &   619.770& 	0.86&	1.18&	1.23 \\
 & 	2449688.987&    15.79 &   619.778& 	0.91&	1.18&	1.24 \\
 & 	2449689.048&    15.79 &   619.783& 	0.94&	1.18&	1.24 \\
\hline\\
%KPNO -- Richards
KPNO2 & 	2449689.767&    15.79 &   619.848& 	0.94&	1.19&	1.29 \\
 & 	2449689.856&    15.80 &   619.856& 	0.93&	1.18&	1.29 \\
 & 	2449689.961&    15.80 &   619.865& 	0.88&	1.19&	1.26 \\
 & 	2449690.055&    15.80 &   619.874& 	0.77&	1.17&	1.24 \\
 & 	2449690.792&    15.80 &   619.940& 	0.65&	1.12&	1.31 \\
 & 	2449690.901&    15.80 &   619.950& 	0.61&	1.11&	1.32 \\
 & 	2449691.003&    15.80 &   619.959& 	0.59&	1.18&	1.31 \\
 & 	2449691.051&    15.80 &   619.963& 	0.58&	1.15&	1.34 \\
 & 	2449694.967&    15.81 &   620.316& 	0.91&	1.24&	1.18 \\
 & 	2449695.048&    15.81 &   620.323& 	0.88&	1.24&	1.19 \\   
 & 	2450129.603&    16.85 &   659.426& 	0.95&	1.23&	1.20 \\
 & 	2450129.641&    16.85 &   659.430& 	0.95&	1.24&	1.20 \\
 & 	2450129.828&    16.85 &   659.447& 	0.94&	1.25&	1.23 \\
 & 	2450130.638&    16.85 &   659.519& 	0.89&	1.21&	1.30 \\
 & 	2450130.747&    16.85 &   659.529& 	0.91&	1.21&	1.29 \\
 & 	2450130.830&    16.85 &   659.537& 	0.91&	1.18&	1.28 \\
 & 	2450130.880&    16.85 &   659.541& 	0.92&	1.15&	1.29 \\
 & 	2450131.624&    16.85 &   659.608& 	1.02&	1.21&	1.28 \\   
 & 	2450131.729&    16.85 &   659.618& 	1.01&	1.22&	1.31 \\
 & 	2450131.828&    16.86 &   659.626& 	1.02&	1.23&	1.29 \\
 & 	2450131.872&    16.86 &   659.630& 	1.01&	1.24&	1.30 \\
 & 	2450132.639&    16.86 &   659.699& 	0.97&	1.26&	1.25 \\
\hline\\
%KPNO -- Peters
KPNO3 & 2451191.906&    19.40 &   755.017&	        0.42&	1.37&	1.25 \\
 & 	2451193.934&    19.40 &   755.199&	        0.67&	1.29&	1.18 \\
 & 	2451196.898&    19.41 &   755.466&	        0.46&	1.19&	1.17 \\   	
 & 	2451502.956&    20.14 &   783.007&	        0.74&	1.47&	1.54 \\
\enddata
\end{deluxetable}

\begin{deluxetable}{lrrrrrr}
\setcounter{table}{1}
\tabletypesize{\scriptsize}
\tablecaption{Optical Observations (continued)}
\tablehead{\colhead{ID\tablenotemark{a}} &\colhead{HJD} & \colhead{${\phi}_{\rm L}$} & 
\colhead{${\phi}_{\rm 0}$} & \colhead{$I_{\rm abs}$} & \colhead{$I_{\rm blue}$} & \colhead{$I_{\rm red}$} }
\startdata
%KPNO -- Peters
KPNO3 & 2451503.997&    20.15 &   783.100&	        1.02&	1.29&	1.22 \\
 &   	2451505.005&    20.15 &   783.191&	        1.10&	1.34&	1.23 \\
 & 	2451510.890&    20.16 &   783.721&	        1.05&	1.31&	1.22 \\
 & 	2451913.685&    21.13 &   819.966&	        0.69&	1.20&	1.21 \\
\hline\\
%HET -- Richards
HET  & 	2452610.824&    22.80 &    882.697&        0.81&    1.24&    1.23 \\
 & 	2452614.898&    22.81 &    883.064&        0.59&    1.31&    1.06 \\
 & 	2452616.812&    22.82 &    883.236&        0.79&    1.26&    1.16 \\
 & 	2452618.823&    22.82 &    883.417&        0.70&    1.26&    1.20 \\
 & 	2452618.906&    22.82 &    883.425&        0.67&    1.25&    1.18 \\
 & 	2452619.817&    22.82 &    883.507&        0.60&    1.22&    1.25 \\
 & 	2452619.906&    22.82 &    883.515&        0.60&    1.20&    1.27 \\
 & 	2452620.898&    22.83 &    883.604&        0.70&    1.23&    1.24 \\
 & 	2452621.888&    22.83 &    883.693&        0.81&    1.25&    1.21 \\
 & 	2452622.882&    22.83 &    883.782&        0.82&    1.16&    1.17 \\
 & 	2452633.879&    22.86 &    884.772&        0.86&    1.13&    1.16 \\
 & 	2452635.869&    22.86 &    884.951&        0.54&    1.10&    1.25 \\
 & 	2452636.858&    22.86 &    885.040&        0.49&    1.24&    1.02 \\
 & 	2452637.754&    22.87 &    885.121&        0.79&    1.23&    1.15 \\
 & 	2452639.772&    22.87 &    885.302&        0.89&    1.20&    1.14 \\
 & 	2452640.765&    22.87 &    885.392&        0.88&    1.22&    1.22 \\
 & 	2452641.771&    22.88 &    885.482&        0.83&    1.22&    1.21 \\
 & 	2452642.851&    22.88 &    885.579&        0.88&    1.15&    1.26 \\
 & 	2452643.750&    22.88 &    885.660&        0.87&    1.23&    1.18 \\
 & 	2452644.754&    22.88 &    885.751&        0.95&    1.17&    1.13 \\
 & 	2452645.764&    22.89 &    885.841&        0.98&    1.17&    1.18 \\
 & 	2452649.739&    22.89 &    886.199&        0.88&    1.22&    1.11 \\
 & 	2452651.804&    22.90 &    886.385&        0.89&    1.25&    1.17 \\
 & 	2452652.710&    22.90 &    886.466&        0.81&    1.22&    1.20 \\
\hline\\
MDM & 2455614.724  & 30.01   &   1163.046  &  0.44  & 1.38   & 1.15   \\ 
    & 2455615.636  & 30.01   &   1163.131  &  0.61  & 1.16   & 1.12   \\
    & 2455617.678  & 30.01   &   1163.312  &  0.87  & 1.14   & 1.06   \\
    & 2455618.651  & 30.02   &   1163.400  &  0.83  & 1.13   & 1.14   \\
    & 2455620.846  & 30.02   &   1163.597  &  0.65  & 1.12   & 1.13   \\
LCO & 2455878.823  & 30.64   &   1186.766  &  0.78  & 1.15   & 1.08 \\                      
\enddata

\tablecomments{The long-period phase ${\phi}_{\rm L}$ is calculated
  from Equation 3 of Desmet et al.~(2010) and the binary photometric
  phase ${\phi}_{\rm 0}$ is calculated from Equation 2 of Desmet et
  al.~(2010) but with the epoch shifted by 1030 periods. The
  quantities $I_{\rm abs}$, $I_{\rm blue}$, and $I_{\rm red}$ provide
  the normalized flux level of the deepest $H\alpha$ central
  absorption and of the maximum H$\alpha$ emission in the blue and red
  wings. }

\tablenotetext{a}{Data sets KPNO1 and KPNO3 were collected by Peters
at KPNO; KPNO2 and HET were collected by Richards at KPNO and HET
respectively; MDM and LCO were collected by Miller and Atwood-Stone.}

\end{deluxetable}

The optical data were reduced using the Image Reduction and Analysis
Facility (IRAF) following standard
methods, briefly outlined below. This description applies to the HET
data; while some specific details necessarily differed across the
various telescopes and instruments, the general process was the same.
Bias frames were averaged and then subtracted from the object and
calibration frames. Flat fields were mode scaled and then median
combined, then the master flat was normalized to remove the blaze
function both along and across the dispersion axis, leaving only
pixel-to-pixel variation; object frames were then divided by the
normalized master flat. 
For each object image, obvious cosmic rays near/in the aperture were identified through inspection and removed through interpolation across columns (where the dispersion is approximately along rows).  Next, the aperture was independently defined and traced for each image, and the one-dimensional spectra were extracted using a variance-weighted sum to minimize noise, with additional cosmic ray and bad pixel rejection performed using $\sigma$ clipping.
The dispersion relation was determined from a ThAr lamp
spectrum obtained near the time of the object exposures. The spectra
were normalized using a Chebyshev function with the H$\alpha$ region
excluded from the fit and the rejection $\sigma$ set significantly
lower for points below the curve (to minimize the influence of sharp
absorption lines). The spectra were corrected to the heliocentric
frame and then to the rest frame of the binary system, assuming the systemic velocity 
$\gamma=17.8$~km~s$^{-1}$ given by Desmet et al.~(2010; their Table 8). 
Telluric atmospheric water vapor lines within the H$\alpha$ region were removed with a model generated from a rapidly rotating standard star.  The model was adjusted to match each observation by scaling the depth of the telluric absorption features to account for changes in airmass or other atmospheric variations; here, the line ratios of the telluric lines near H$\alpha$ were assumed to be constant.  When necessary, minor shifts in the model wavelengths were also made.  Finally, any remaining obvious instrumental or cosmic ray artifacts were corrected by interpolating across neighboring non-affected pixels. 

%% Figure 2 HERE

\begin{figure*}
\figurenum{2}
\epsscale{1.1}
\plotone{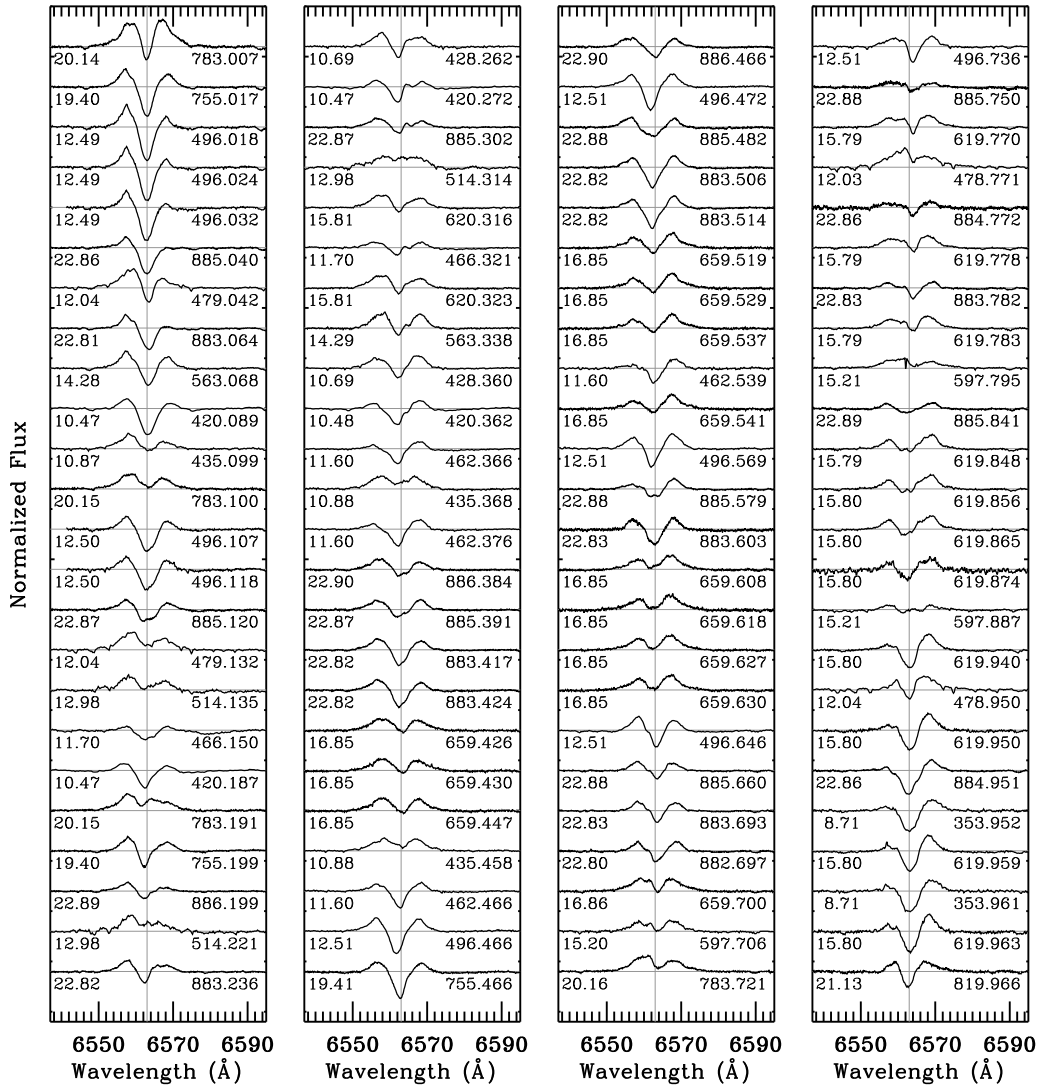}
\vspace{-1.5cm}
\caption{Normalized H$\alpha$ spectra of AU Mon. The long-period phase
  ${\phi}_{\rm L}$ and the binary photometric phase ${\phi}_{\rm 0}$
  are indicated at left and right, respectively, with the epoch also
  provided prior to the decimal. The spectra are shown in order of
  increasing ${\phi}_{\rm 0}$. The HET/HRS data span ${\phi}_{\rm
    0}=882-886$.}
\label{f2}
\end{figure*}

We also analyzed forty-three (43) UV spectra of AU Mon that were
collected with the {\it International Ultraviolet Explorer\/} ({\it
IUE\/}) on various dates over 16 years from 1978 to 1994. Only the
short-wavelength high-resolution {\it IUE\/} spectra are considered
here; these observations include the most useful diagnostic lines 
(\ion{Si}{4}, \ion{Si}{2}, \ion{Al}{3}, etc.) and are of sufficient 
resolution to permit a detailed comparison with the {\it Shellspec\/} models.
The SWP camera covers a range from 1150--1900~\AA~with a
resolution of $\sim$10,000. These data have been previously presented,
analyzed, and discussed by Peters \& Polidan (1982), Peters \& Polidan
(1984), and Peters (1994). In this paper, we extend the earlier work by
utilizing these UV spectra along with {\it Shellspec\/} modeling to
ascertain and confirm the parameters for the accretion disk and gas
stream of AU Mon.

The binary photometric phase ${\phi}_{\rm 0}$ for each observation was
calculated with the ephemeris given by Desmet et al.~(2010), except
that the HJD$_{\rm min}$ given in their Equation 2 has been modified
by subtracting 1030 periods so that all the epochs are positive (the
effective HJD$_{\rm min}$ then nearly matches that given in Table 2 of
Richards \& Albright~1999). Figure 1 shows the dense phase coverage
obtained of AU Mon at both optical and UV wavelengths. The long-period
phase ${\phi}_{\rm L}$ for each observation was calculated from
Equation 3 of Desmet et al.~(2010).  While we label specific
observations as {\it epoch.phase\/}, we discuss general properties
using $0\le\phi<1$ notation throughout. Some basic properties of the
optical and UV data are given in Tables 1 and 2, respectively.
 
%% Figure 3 HERE

\begin{figure*}
\figurenum{3}
\epsscale{1.18}
\plotone{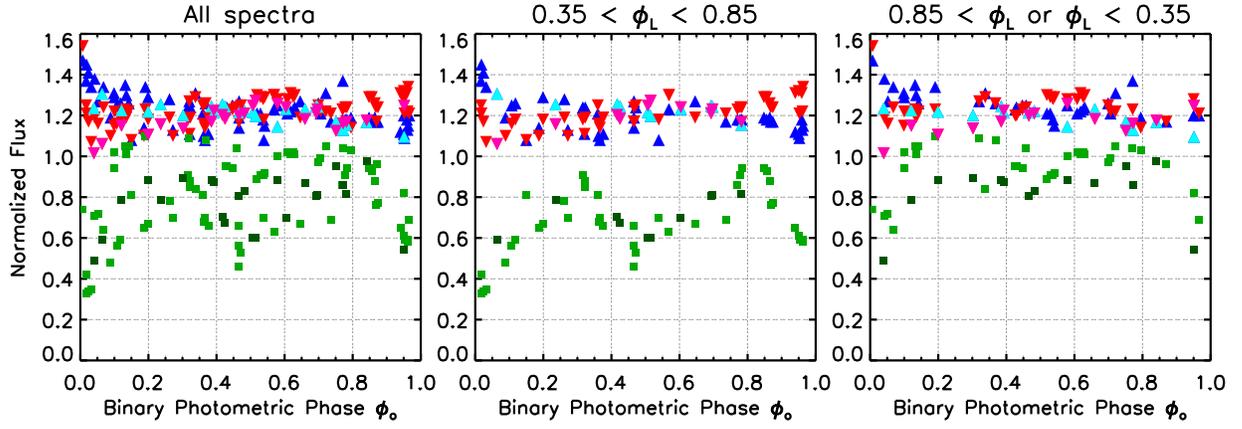}
\vspace{0.01cm}
\caption{A REBECA diagram showing the Red Emission (red triangles),
  Blue Emission (blue triangles), and Central Absorption (green
  squares) for AU Mon versus the binary photometric phase ${\phi}_{\rm
    0}$. The data points indicate the maximum normalized flux for the
  emission wings and the deepest point for the central absorption. The
  HET/HRS data are distinguished with slightly modified colors
  (pink/cyan/dark green). When the data are separated by long-period
  phase, it is apparent that the central absorption is deeper for
  $0.35<{\phi}_{\rm L}<0.85$.}
\label{f3}
\end{figure*}

%%TABLE 2 HERE
\begin{deluxetable*}{lrrrrrrrrrr}
\tabletypesize{\scriptsize}
%\rotate
\tablecaption{UV Observations}
\tablewidth{15.5cm}

\tablehead{\colhead{HJD} & \colhead{${\phi}_{\rm L}$} &
  \colhead{${\phi}_{\rm 0}$} & \multicolumn{2}{c}{\ion{Si}{4} $EW$
    (\AA)} & \multicolumn{2}{c}{\ion{Si}{2} $EW$ (\AA)} &
  \multicolumn{2}{c}{\ion{Al}{3} $EW$ (\AA)} &
  \multicolumn{2}{c}{\ion{Al}{3} $v$ (km/s)\tablenotemark{a}} \\ & & &
  \colhead{$\lambda$1394} & \colhead{$\lambda$1403} &
  \colhead{$\lambda$1264} & \colhead{$\lambda$1309} &
  \colhead{$\lambda$1855} & \colhead{$\lambda$1863} &
  \colhead{$\lambda$1855} & \colhead{$\lambda$1863}}

\startdata

2443873.902 &    1.87 &   96.50 & 1.43 & 1.08 & 0.50 & 0.39 & 1.14 & 0.85 & -20.5 &   5.6 \\ 
2443875.497 &    1.87 &   96.64 & 1.49 & 1.19 & 0.83 & 0.45 & 1.51 & 1.19 &   8.0 &  25.9 \\ 
2444484.907 &    3.36 &  151.48 & 1.12 & 0.92 & 0.93 & 0.72 & 1.34 & 1.15 &  25.6 &  31.1 \\ 
2444510.971 &    3.42 &  153.83 & 1.30 & 0.94 & 0.91 & 0.65 & 1.04 & 0.93 &  83.8 &  92.3 \\ 
2444612.260 &    3.67 &  162.94 & 1.07 & 0.88 & 0.94 & 0.73 & 1.27 & 1.16 &  46.3 &  57.1 \\ 
2444616.365 &    3.68 &  163.31 & 0.81 & 0.59 & 1.25 & 0.65 & 1.04 & 0.79 &   9.7 &   8.5 \\ 
2445800.515 &    6.56 &  269.86 & 1.52 & 1.14 & 1.48 & 0.84 & 1.20 & 1.14 &  61.7 &  71.0 \\ 
2445977.926 &    6.99 &  285.83 & 1.61 & 1.26 & 0.81 & 0.57 & 1.41 & 1.23 &  42.1 &  79.1 \\ 
2445982.788 &    7.00 &  286.27 & 1.14 & 0.98 & 0.94 & 0.49 & 1.13 & 0.82 & -72.3 & -35.6 \\ 
2447422.844 &   10.51 &  415.85 & 1.11 & 0.81 & 1.25 & 0.94 & 1.06 & 1.01 &  91.0 &  98.3 \\ 
2447427.074 &   10.52 &  416.23 & 0.87 & 0.60 & 0.86 & 0.65 & 0.83 & 0.82 &  -9.6 &   0.9 \\ 
2447550.515 &   10.82 &  427.34 & 1.07 & 0.79 & 0.93 & 0.67 & 1.25 & 1.06 &  15.1 &  24.2 \\ 
2447556.534 &   10.83 &  427.88 & 1.24 & 0.96 & 0.96 & 0.63 & 1.14 & 1.00 &  57.3 &  67.0 \\ 
2447867.624 &   11.59 &  455.87 & 0.93 & 0.74 & 1.35 & 0.85 & 1.01 & 0.94 &  74.5 &  82.5 \\ 
2447789.852 &   11.40 &  448.87 & 1.31 & 1.12 & 1.02 & 0.61 & 1.19 & 1.14 & 107.2 & 133.1 \\ 
2447794.882 &   11.41 &  449.33 & 0.82 & 0.76 & 0.75 & 0.55 & 0.95 & 0.90 &   2.7 &   7.8 \\ 
2447871.795 &   11.60 &  456.25 & 0.97 & 0.75 & 1.14 & 0.63 & 0.96 & 0.92 &  19.3 &  25.8 \\ 
2447938.370 &   11.76 &  462.24 & 1.10 & 0.73 & 1.33 & 0.69 & 0.87 & 0.79 & -27.4 & -15.2 \\ 
2447945.359 &   11.78 &  462.87 & 1.44 & 0.80 & 1.24 & 0.81 & 1.10 & 1.05 &  62.8 &  60.6 \\ 
2448001.258 &   11.91 &  467.90 & 1.27 & 1.03 & 0.78 & 0.41 & 0.90 & 0.94 &  67.7 &  67.7 \\ 
2448004.453 &   11.92 &  468.18 & 1.31 & 0.76 & 0.91 & 0.58 & 0.87 & 0.88 &  16.6 &  21.4 \\ 
2448228.619 &   12.47 &  488.36 & 0.90 & 0.73 & 0.92 & 0.62 & 0.92 & 0.93 &  27.8 &  24.8 \\ 
2448229.691 &   12.47 &  488.45 & 1.05 & 0.87 & 0.82 & 0.62 & 1.10 & 1.05 &  14.7 &  36.5 \\ 
2448230.606 &   12.47 &  488.53 & 1.29 & 0.92 & 1.17 & 0.53 & 1.30 & 1.14 &  26.5 &  29.9 \\ 
2448231.610 &   12.47 &  488.63 & 1.56 & 1.24 & 1.03 & 0.63 & 1.49 & 1.34 &  30.5 &  38.2 \\ 
2448234.611 &   12.48 &  488.90 & 1.75 & 1.21 & 1.31 & 0.65 & 1.39 & 1.24 &  52.4 &  76.4 \\ 
2448349.469 &   12.76 &  499.23 & 1.00 & 0.64 & 0.99 & 0.65 & 0.79 & 0.81 &   3.1 &   1.2 \\ 
2448351.472 &   12.77 &  499.41 & 0.94 & 0.70 & 1.07 & 0.72 & 1.00 & 0.93 &  12.5 &   2.9 \\ 
2448352.460 &   12.77 &  499.50 & 1.00 & 0.77 & 1.18 & 0.77 & 0.96 & 0.98 &  15.4 &  18.0 \\ 
2448353.509 &   12.77 &  499.59 & 0.99 & 0.84 & 1.05 & 0.76 & 1.16 & 1.06 &  37.5 &  39.2 \\ 
2448356.468 &   12.78 &  499.86 & 1.19 & 0.86 & 1.83 & 1.03 & 1.25 & 1.22 &  59.1 &  60.6 \\ 
2448912.776 &   14.13 &  549.92 & 1.76 & 1.33 & 1.27 & 0.42 & 1.35 & 1.26 &  96.0 & 122.6 \\ 
2448915.998 &   14.14 &  550.21 & 1.41 & 0.94 & 0.79 & 0.48 & 0.86 & 0.94 & -29.7 & -32.0 \\ 
2449335.627 &   15.16 &  587.97 & 1.21 & 1.13 & 0.84 & 0.61 & 1.34 & 1.31 &  99.9 & 126.3 \\ 
2449375.399 &   15.26 &  591.55 & 2.07 & 1.04 & 0.79 & 0.42 & 1.54 & 1.17 & -33.1 &   2.9 \\ 
2449379.416 &   15.27 &  591.91 & 1.68 & 1.37 & 1.04 & 0.49 & 1.50 & 1.54 &  89.0 & 113.0 \\ 
2449383.412 &   15.28 &  592.27 & 1.93 & 1.31 & 0.85 & 0.50 & 0.75 & 0.77 & -68.4 & -76.6 \\ 
2449658.929 &   15.95 &  617.06 & 1.35 & 0.97 & 0.92 & 0.60 & 1.06 & 0.87 &  12.7 &  21.0 \\ 
2449690.597 &   16.02 &  619.91 & 1.29 & 1.06 & 0.86 & 0.50 & 1.10 & 1.05 &  27.9 &  36.9 \\ 
2449691.321 &   16.03 &  619.98 & 1.22 & 0.74 & 1.04 & 0.57 & 0.95 & 1.04 &  37.8 &  57.4 \\ 
2449691.787 &   16.03 &  620.02 & 1.78 & 1.37 & 0.93 & 0.58 & 1.20 & 1.26 &   8.9 &  31.5 \\ 
2449694.603 &   16.03 &  620.27 & 1.12 & 0.73 & 0.95 & 0.50 & 0.81 & 0.82 &  13.0 &  19.8 \\ 
2449696.789 &   16.04 &  620.47 & 1.72 & 1.37 & 0.97 & 0.46 & 1.54 & 1.37 &  -5.5 &  -0.5 \\ 
\enddata

\tablecomments{The long-period phase ${\phi}_{\rm L}$ and the binary
  photometric phase ${\phi}_{\rm 0}$ are calculated as in Table 1.}

\tablenotetext{a}{The velocity is calculated from the absorption-weighted line center.}

\end{deluxetable*}

The optical H$\alpha$ spectra are shown in Figure 2. The
phase-dependent properties of the H$\alpha$ line profile are
illustrated (Figure 3) in a ``REBECA'' diagram [Red Emission, Blue
Emission, Central Absorption; see also Budaj et al.~(2005) and Miller
et al.~(2007) for discussion] showing the maximum normalized flux in
each emission wing (red and blue triangles) and the deepest point of
absorption (green squares). AU Mon shows many of the general
characteristics also apparent in TT Hya (above references), including
partial disk eclipses near primary eclipse and enhanced central
absorption near primary eclipse and to a lesser degree near secondary
eclipse. However, there is significantly more scatter in the REBECA
diagram of AU Mon, particularly in the central absorption. This
scatter makes it difficult to ascertain whether additional effects
seen in TT Hya are also present in AU Mon; for example, it appears
that the tendency for greater blue emission at ${\phi}_{\rm
0}=0.4-0.5$ and greater red emission at ${\phi}_{\rm 0}=0.5-0.6$
(thought to arise from absorption of the light from the secondary by
one side of the disk as well as by the primary; above references) is
suggested but not required by these data. The scatter in the central
absorption is reduced when the spectra obtained over different regions
of the long-period phase are considered separately, and those taken
within $0.35<{\phi}_{\rm L}<0.85$ show deeper central absorption than
those taken within ${\phi}_{\rm L}>0.85$ or ${\phi}_{\rm
L}<0.35$. Similar tendencies have been identified by Barr{\'{\i}}a \&
Mennickent~(2011).

%% Figure 4 HERE

\begin{figure*}
\figurenum{4}
\epsscale{1.0}
\plotone{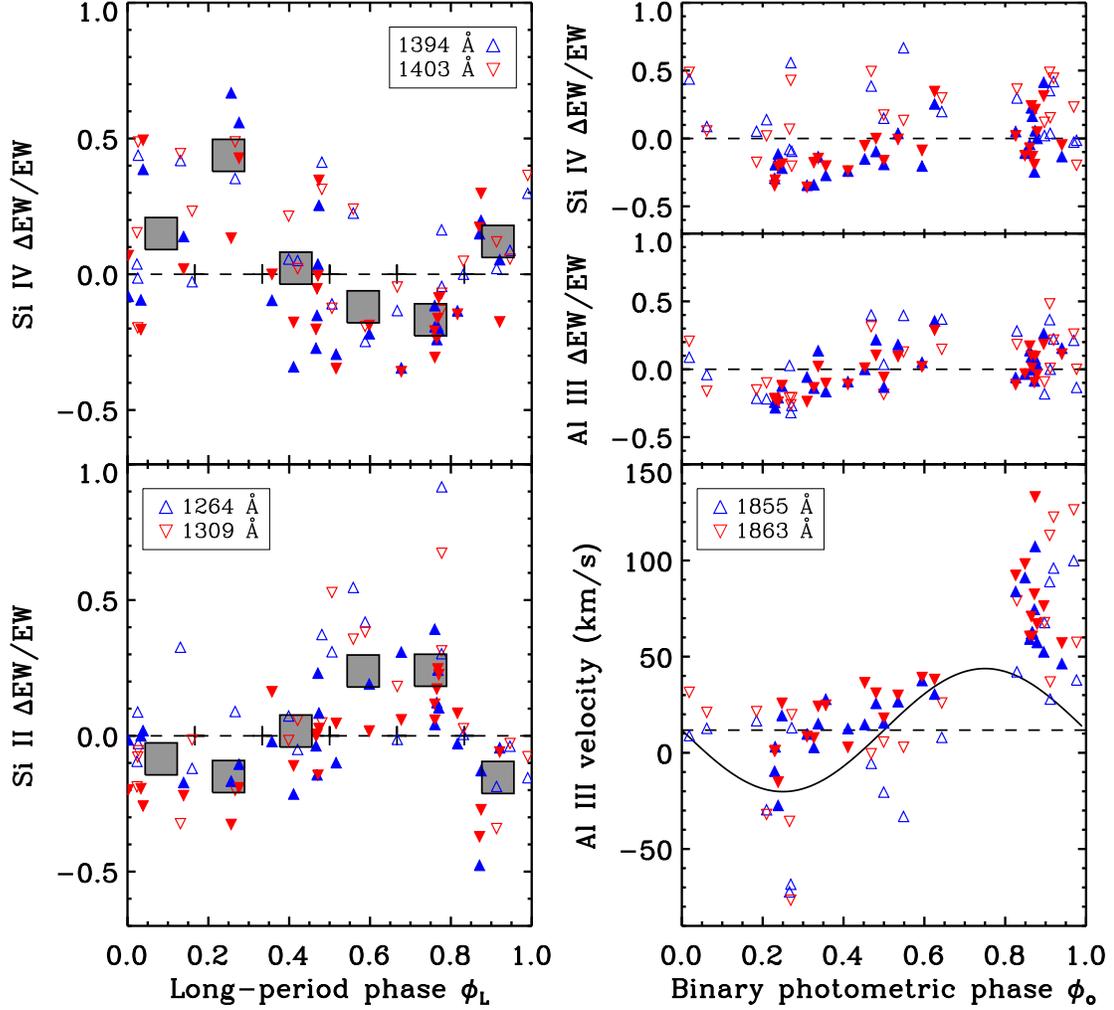}
\vspace{-1cm}
\caption{Properties of the UV absorption features in AU Mon plotted
  versus long-period phase ${\phi}_{\rm L}$ (left column) and binary
  photometric phase ${\phi}_{\rm 0}$ (right column). The filled points
  in the left-hand column are for $0.1<{\phi}_{\rm 0}<0.8$ (i.e.,
  outside of primary eclipse and outside of the absorption influence
  of the gas stream), and the filled points in the right-hand column
  are for $0.35<{\phi}_{\rm L}<0.85$. The fractional change in
  equivalent width ${\Delta}EW/EW$ is calculated for each point as
  $(EW-EW_{\rm med})/EW_{\rm med}$, where $EW_{\rm med}$ is the median
  equivalent width. The gray squares are means of all points within
  that sixth of the phase cycle (bounds indicated on the central
  dashed line). Velocities are for the absorption-weighted line
  center.}
\label{f4}
\end{figure*}

The UV line profiles also show phase-dependent variation, as
illustrated in Figure 4. The fractional change in equivalent width is
calculated for each observation as $(EW-EW_{\rm med})/EW_{\rm med}$,
where $EW_{\rm med}$ is the median equivalent width, and the velocity
is calculated from the offset of the absorption-weighted line center
from the lab wavelength for that transition. 
The relative absorption strength of \ion{Si}{4} versus \ion{Si}{2} shows a striking dependence on the long-period phase, which reflects changes in the disk absorption, gas stream variability, and overall heating of the photosphere (Peters 1994). The long-term variation in Si II is primarily due to changes in disk absorption that maximizes around a ${\phi}_{\rm L}$ of $0.5-0.8$. The Si IV variations are predominately on the trailing hemisphere and reflect an enhancement of the HTAR (Peters \& Polidan 1984) and some Si IV absorption from the gas stream. The correlation with ${\phi}_{\rm L}$ of H$\alpha$ central absorption, \ion{Si}{4} and other HTAR lines {\it versus} \ion{Si}{2} equivalent widths and the UV flux distribution do not support ascribing the long-term brightness variation to attenuation by circumbinary material (Desmet et al.~2010; Djura{\v s}evi{\'c} et al.~2010). Instead, these periodic tendencies are consistent with the hypothesis of variable mass transfer over the long-period phase (e.g., due to oscillations in the size of the secondary), peaking near ${\phi}_{\rm L}\sim0.4$, with the consequent build up of a thick disk over ${\phi}_{\rm L}=0.5-0.8$ (Peters 1994; the long-period phases in that paper should be adjusted by about $-0.1$ to match the ephemeris of Desmet et al.~2010). The circumprimary envelope conjectured by Barr{\'{\i}}a \& Mennickent~(2011) is a similar concept.  A thick disk, enduring over 10--20 orbital periods, can produce both the enhanced \ion{Si}{2} absorption and the deeper H$\alpha$ central absorption seen during $0.35<{\phi}_{\rm L}<0.85$, and typically decrease the light from the primary through continuous and line opacities. The collapse of the thick disk could then heat the photosphere of the  B star and perhaps generate the 10000--14000~K plasma suggested by Peters (1994) as the well-tuned system recovers to equilibrium before the next episode of enhanced mass transfer.  In this scenario, the self-similar orbital light curve shape at long-period maxima and minima (Desmet et al.~2010; Djura{\v s}evi{\'c} et al.~2010) may indicate that as the thick disk collapses the outer regions puff up or flare to extend beyond the orbital plane, which is potentially related to the L3 mass loss near maximum long-period brightness proposed by Peters (1994). 

Some of the UV line profiles also
vary with the binary photometric phase ${\phi}_{\rm 0}$. For example,
the \ion{Al}{3} doublet shows increased redshifted absorption near the
binary photometric phase of ${\phi}_{\rm 0}\sim0.85$, ascribed by
Peters (1994) to absorption within the infalling gas stream. In
$\S$3.2, we demonstrate through {\it Shellspec\/} modeling that this
effect can indeed be produced by the gas stream.

%%%%%%%%%%%%%%%%%%%%%%%%%%%%%%%%%%%%%%%%%%%%%%%%%%%%%%%%%%%%%

\section{Shellspec Modeling}

The system parameters used to create the synthetic spectra were primarily 
taken from the mean CoRoT disk model
of Djura{\v s}evi{\' c} et al.~(2010; their Table~1). In particular,
we adopt their system inclination of 80.1 degrees, component
separation of 42.1~$R_{\odot}$, primary mass of 7.0~$M_{\odot}$, mass
ratio of $q=0.17$, and primary radius of 5.1~$R_{\odot}$; while the
secondary dimensions are specified by the Roche surface calculated
within {\it Shellspec\/} (Budaj \& Richards~2004). We briefly investigated alternative values
of inclination and primary radius in our modeling and confirmed that
the Djura{\v s}evi{\' c} et al.~(2010) values gave optimal
results. The primary and secondary velocities are set to 30.9 and 161
km~s$^{-1}$, respectively, which are deprojected from the measured $K$
values given in Desmet et al.~(2010; their Table~6). The system
parameters are fixed throughout the process of modeling the accretion
disk and gas stream.

%%TABLE 3 HERE

\begin{deluxetable*}{lrrrr}
%\tabletypesize{\scriptsize}
%\rotate
\tablecaption{Shellspec Model Parameters}
\tablewidth{14cm}

\tablehead{\colhead{Parameter} & \colhead{Primary} &
\colhead{Secondary} & \colhead{Accretion disk} & \colhead{Gas stream}}

\startdata

Temperature (K)                                 & 17000   & 5750    & 14000\tablenotemark{a}   & 8000     \\
Radius\tablenotemark{b} ($R_{\odot}$)           & 5.1     & 10.1    & 5.1, 23 & 4.0      \\
Disk thickness ($R_{\odot}$)                    & \nodata & \nodata & 5.2     & \nodata  \\
Mass ($M_{\odot}$)                              & 7.0     & 1.2     & \nodata & \nodata  \\
Density (10$^{-15}$ g cm$^{-3}$)                & \nodata & \nodata & 100     & 60       \\
$v_{\rm rot}$ (km s$^{-1}$)                     & 118.3   & 44.7    & \nodata & \nodata  \\
$v_{\rm turb}$ (km s$^{-1}$)                    & \nodata & \nodata & 30      & 100      \\
$v_{\rm orb}$ (km s$^{-1}$)                     & $-$30.9 & 161     & $-$30.9 & \nodata  \\
$v_{\rm init}$, $v_{\rm final}$ (km s$^{-1}$)   & \nodata & \nodata & \nodata & 100, 500 \\
\enddata

\tablecomments{The modeling process and the estimated uncertainties
are detailed in $\S$3.}

\tablenotetext{a}{The temperature of the accretion disk is a function
of radius; the given value is the maximum temperature.}

\tablenotetext{b}{The radius of the secondary is the average Roche
lobe radius. The inner and outer disk radii are given. The separation
between the primary and secondary is 42.1 $R_{\odot}$.}

\end{deluxetable*}

The general method used to produce and evaluate {\it Shellspec\/}
models was as follows. First, synthetic stellar spectra were generated
for the primary and secondary stars, using the program SPECTRUM (Gray
\& Corbally 1994). SPECTRUM was run with model stellar atmospheres
produced by Castelli \& Kurucz~(2003) with ATLAS9. The effective
temperature and surface gravity for the primary were identified
through comparison to synthetic spectra with particular attention paid
to the UV lines; we find that a B3 stellar type with $T\simeq17000$ K
and $\log{g}\simeq3.5$ provides a good match. Note that this
temperature is more consistent with the primary mass found by Desmet
et al.~(2010), whereas they note that their preferred value is
somewhat lower than is typical for this spectral class. For the
secondary, we assume $T=5750$ K and $\log{g}=2.5$, as found by Desmet et
al.~(2010). The resulting synthetic stellar spectra were used as input
for {\it Shellspec\/}, which was then run to create synthetic
composite spectra for the system at phases matching our observational
coverage. The parameters specifying the primary and secondary star,
given in Table~3, were held fixed throughout the modeling process and
can be compared with the separation between the primary and secondary,
which is 42.1 $R_{\odot}$.  The shape of the secondary star is
calculated by {\it Shellspec\/} assuming that it fills its Roche lobe, hence the
quoted radius is the average Roche lobe radius. The parameters
specifying the accretion disk were set interactively for each trial;
Table~3 lists those from our final preferred model. A gas stream (with
parameters held constant) was also included for most of the modeling, as
described below. Finally, the output synthetic composite spectra were
compared to the observed spectra at a range of phases. As part of this
process, difference profiles were computed by subtracting the
synthetic from the observed spectra.

We identified through trial and error an initial set of accretion disk
parameters that provided a reasonable match to the observed optical
and UV spectra, then considered the effect of the mass-transfer stream
before further refining the disk model. The influence of the gas
stream is subtle relative to that of the disk, and the precise
properties of the stream are only loosely constrained here. After
examining a range of parameters, we adopted a set that successfully
reproduces the redshifted absorption observed in \ion{Al}{3} (and to a
lesser degree in H$\alpha$) at $0.75<{\phi}_{\rm 0}<1.0$. This stream
component was then included unchanged in all subsequent modeling,
while the disk parameters were varied over a grid to improve the disk
model; consistent inclusion of the relatively minor contribution of the gas
stream throughout reduces the possibility of biasing the disk
properties by attempting to match an incomplete model to the
observations. Additional iterations (e.g., testing a grid of stream
models around the best disk model) did not provide additional
information.

\subsection{The Accretion Disk}

The free parameters of interest for the accretion disk are the
thickness, the outer radius, the characteristic temperature, and the
density. The inner disk radius is fixed to be slightly larger than the
radius of the primary star, at 5.2 $R_{\odot}$. The temperature of the
disk is a function of radius as given by Equation~58 of Budaj \&
Richards (2004); e.g., for a characteristic temperature of 27000~K,
the disk temperature increases to a maximum value of $\simeq$13200~K
at a radius of 9 $R_{\odot}$ and then decreases to a temperature of
$\simeq$9500~K at 20 $R_{\odot}$. The density of the disk is set to
scale inversely with radius, with the initial density specified at the
inner radius of the disk. 

A total of 625 potential models were generated by varying these four
parameters through five different values each, straddling the
parameters of the initial model used to characterize the
stream. Specifically, the considered values spanned
2.8--7.6 $R_{\odot}$ in steps of 1.2 $R_{\odot}$ for thickness;
17--25 $R_{\odot}$ in steps of 2 $R_{\odot}$ for outer radius;
25000--33000~K in steps of 2000~K for characteristic temperature; and
70--130$\times10^{-15}$~g~cm$^{-3}$ in steps of
15$\times10^{-15}$~g~cm$^{-3}$ for density. The degree to which each
model matched the data was assessed through calculation of a
goodness-of-fit statistic (sum of squared residuals within the
relevant spectral region), determined for each model at each of the 24
HET phases for H$\alpha$ and H$\beta$ and at each of the 47 {\it
IUE\/} phases for \ion{Al}{3} and \ion{Si}{4}. The other H$\alpha$
data were not considered in this process, since it is desirable to model
the H$\alpha$ emission over a sufficiently short time interval to
exclude long-term dynamical changes that might occur in the accretion
disk.

%%FIGURE 5 HERE

\begin{figure}[t]
\figurenum{5}
\epsscale{1.2}
\plotone{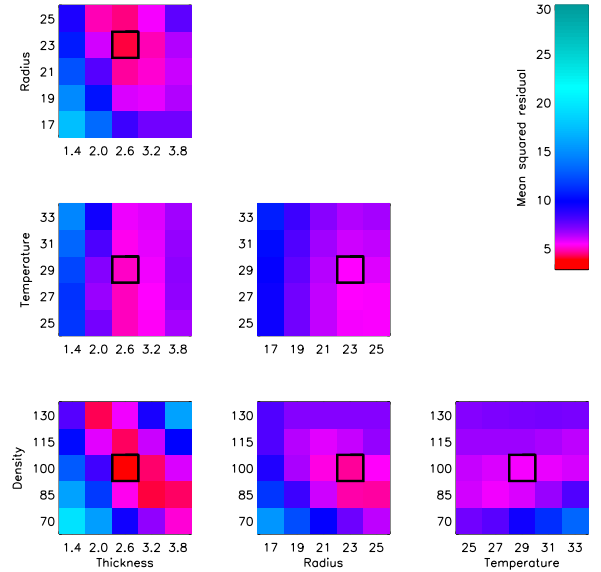}
\vspace{-1cm}
\caption{Accretion disk modeling, evaluated against HET H$\alpha$
spectra taken at a representative set of nine phases. The accretion
disk parameters varied are the disk half-thickness, the outer disk
radius, the effective temperature, and the density; each parameter is
stepped through five values for a total of 625 considered models. The
gas stream is included in the modeling with fixed parameters. Models
are compared using a goodness-of-fit measure (summed squared
residuals), which is shown color-coded (key on right) for projections,
by mean, onto parameter subgrids. The overall preferred model is
indicated by the bold-bordered squares.}
\label{f5}
\end{figure}

The results of the modeling are shown in Figure~5 for a representative
set of nine HET/H$\alpha$ phases covering the full orbit. The
agreement with the fit is illustrated using a color scale (red
indicates good agreement, blue a poor match) for parameter pairs; the
goodness-of-fit indicator is projected across the other two parameters
using mean values. Each cell is therefore constructed from nine phases
and projected down from 25 models. While there is some unavoidable
degeneracy in the H$\alpha$ modeling (e.g., increased disk emission
may be obtained either from increasing the density or the thickness),
there are clearly favored regions of parameter space. Specific
consideration of the eclipse phases is helpful for clarifying the
structure, and the goodness-of-fit for \ion{Al}{3} and \ion{Si}{4}
across identical grids provides useful additional constraints upon
temperature. It appears that models at the extreme ends of the grid
are excluded; for example, there is only poor agreement with
observations for models with simultaneously large values of thickness
and outer radius, or thickness and density, or outer radius and
density, and similarly poor agreement for simultaneously small values
of those parameters. The disk is most readily apparent in the
double-peaked H$\alpha$ emission, and hence the HET/H$\alpha$ data
drive the modeling. However, the disk is also visible in the
difference profiles for other optical (e.g., H$\beta$) and UV lines, 
and the additional spectral constraints help lift
degeneracies in the disk parameters that are present if analysis is
restricted to the H$\alpha$ line.

From the {\it Shellspec\/} modeling, we derived preferred parameters
for the accretion disk (marked on Figure~5 by bold-bordered squares; see also Table~3): a
thickness of 5.2 $R_{\odot}$, an outer radius of 23 $R_{\odot}$, a
characteristic temperature of 29000~K (corresponding to a maximum
temperature of $\simeq$14000~K), and a density of
$100\times10^{-15}$~g~cm$^{-3}$, with a turbulence velocity of
30~km~s$^{-1}$. Some of the other disk models considered cannot be
ruled out; for example, with higher densities
($130\times10^{-15}$~g~cm$^{-3}$) a smaller thickness (4.0 $R_{\odot}$)
is possible, or alternatively with lower densities
($85\times10^{-15}$~g~cm$^{-3}$) a larger thickness (6.4 $R_{\odot}$)
is possible. A larger outer disk radius is tenable, and in general
models bordering the preferred parameters acceptably match the
data.

%%FIGURE 6 HERE

\begin{figure*}
\vspace{-0.6cm}
\figurenum{6}
\epsscale{1.0}
\plotone{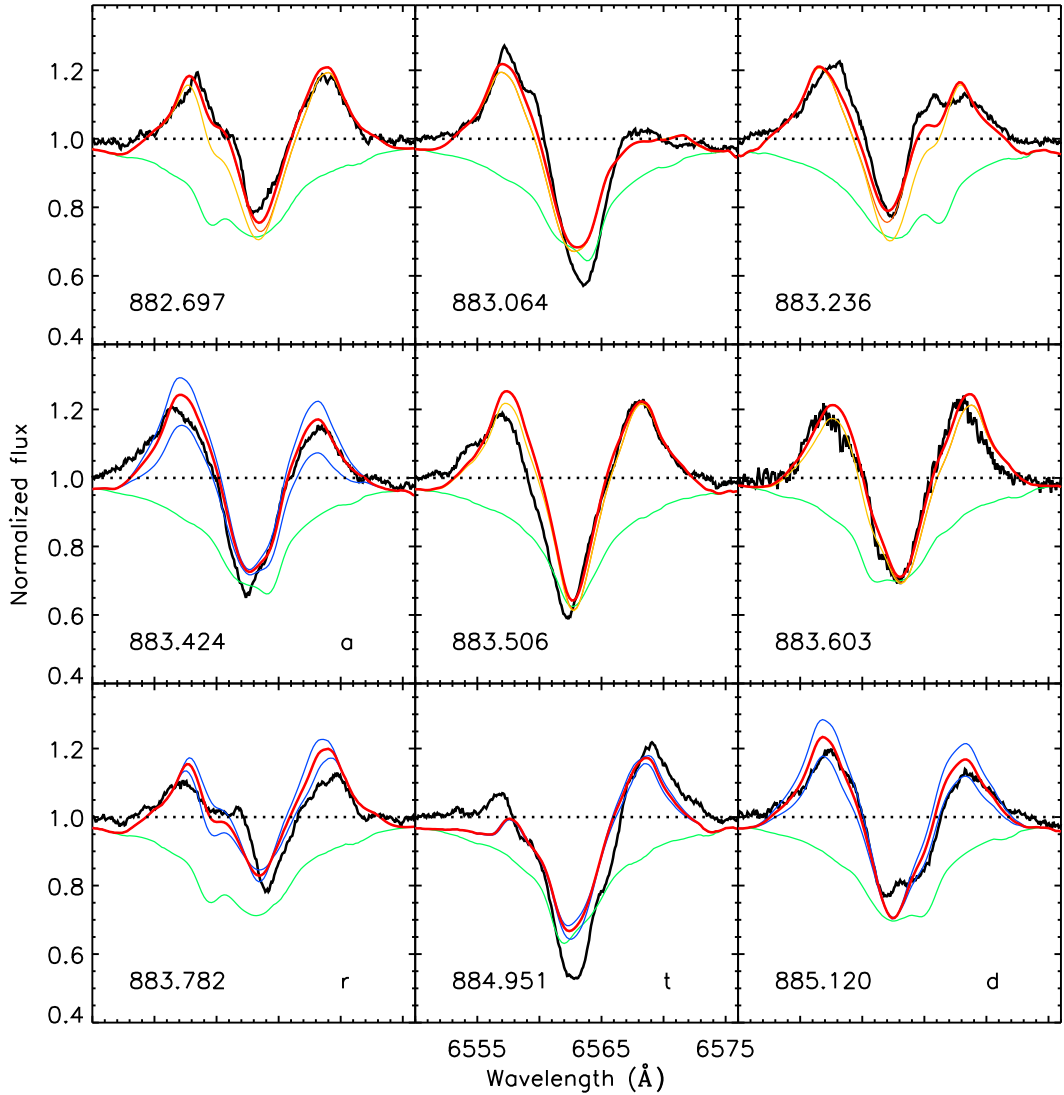}
\caption{The preferred model for the accretion disk+stream+spot is plotted
(red) along with the HET H$\alpha$ profiles (black) at the nine
considered representative phases; the stars-only model is shown in
green for reference. The four frames at the bottom and middle left show (in blue) the
results of changing the disk half-thickness, outer radius, effective temperature, and
density by $\pm$1 step along the grid displayed in Figure~5. The five frames at the top and middle right show the stars+disk and stars+disk+stream models in yellow and orange, respectively.}
\label{f6}
\end{figure*}

By examining the H$\alpha$ profiles (e.g., Figure~6), we found
that the preferred disk plus stream model gives a slightly improved
match to observations with the additional inclusion of a relatively
minor ``spot'' component. This spot is located at the outer disk edge
at coordinates of ($-$18, $-$18) $R_{\odot}$ (in a primary-centered
system) with a radius of 2.6 $R_{\odot}$, a temperature of 8000~K, a
density of 80$\times$10$^{-15}$~g~cm$^{-3}$, a velocity vector of
($-$120, 120)~km~s$^{-1}$, and a turbulence of 100~km~s$^{-1}$. These
parameters have large uncertainties of $\sim$30--40\%, since this spot
has only an incremental impact on the model profile. This ``spot'' may be
physically associated with disk asymmetries, such as caused by an
elliptical rather than circular shape, a clumpy structure, or brighter
regions.  One example is the high density Localized Region along the line of centers between 
the stars where the gas that has circled the primary slows down because of contact with the incoming gas stream (Richards 1992). 

%%FIGURE 7 HERE

\begin{figure}[h]
\figurenum{7}
\epsscale{1.0}
\plotone{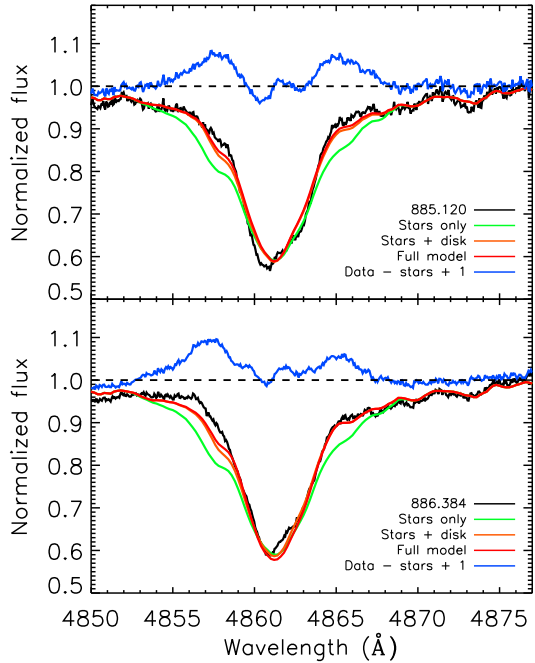}
\vspace{-0.5cm}
\caption{Model for the H$\beta$ line compared to HET spectra (black) at ${\phi}_{\rm 0}=0.120$ (top) and ${\phi}_{\rm 0}=0.384$ (bottom). The model with only stars (green) has deeper absorption wings than are observed. Adding an accretion disk (orange; see $\S$3.1) significantly improves the match, and the full model (red) provides reasonable agreement to the data. The difference profile (blue) shows the underlying double-peaked emission signature of the disk, similar to what is seen directly in the observed H$\alpha$ profile.  The synthetic stellar spectrum for the primary was calculated with $\log{g}=3.0$ only in the case of H$\beta$ to provide a better match with the observed profile.
}
\label{f7}
\end{figure}

The preferred model (including stars, disk, stream, and spot) is plotted in Figure~6 along with the same representative set of nine HET/H$\alpha$ spectra used to generate
Figure~5. In the lower left subplots, the effect of varying one
parameter in isolation by one step along the grid is shown. Our
estimated uncertainties on the modeled accretion disk parameters are
one grid step; for example, the outer radius is
23$\pm$2 $R_{\odot}$. In the upper right subplots, the models for the
disk alone and for the disk plus stream are also shown. 
In the model, it is clear that a substantial fraction of the H$\alpha$ emission originates from the disk.  During the partial primary eclipse, only a small fraction of the central disk region is eclipsed, which may slightly reduce the H$\alpha$ emission. However, since the contribution to the spectral continuum due to the primary star (the dominant source of light in the system) is significantly diminished at these phases, the H$\alpha$ emission is strongly enhanced relative to that continuum. This result occurs because the spectra are not displayed in absolute units but are normalized relative to the continuum.  This effect is naturally taken into account in {\it Shellspec\/}.

Our modeling suggests that the accretion disk extends to larger radii
and is thicker than found by Djura{\v s}evi{\'c} et al.~(2010) from
their analysis of CoRoT and $V$-band light curve data. However, our
{\it Shellspec\/} calculations are primarily based on H$\alpha$
spectroscopy for which the disk is visible in emission even when the
line is optically thin; in contrast, the disk as modeled by Djura{\v
s}evi{\'c} et al.~(2010) is non-transparent (optically thick in the
continuum), which results in smaller dimensions. In addition, their
disk includes a hot spot and two bright spots, which produce
embedded compact regions of enhanced emission, while our disk is
circularly symmetric. Consequently, the differences in disk dimensions
are not necessarily in conflict. Indeed, our disk may be regarded as
having a wavelength-dependent effective size, which is a function of
the opacities in the lines and in the continuum. It is additionally
possible that the parameters of the disk when modeled by Djura{\v
s}evi{\'c} et al.~(2010) physically differed from those present when
we collected our HET observations, which serve as the primary basis for our
modeling. The large disk in AU~Mon may explain the particularly prominent H$\alpha$ emission that is visible near primary eclipse; this is different from V393 Sco, for which the disk is likely to be more optically thick and less extended (Mennickent et al.~2012).

\subsection{The Gas Stream}

The free parameters of interest for the gas stream are the initial and
final velocities, the initial density, the temperature, the radius,
and the turbulence.  In the {\it Shellspec\/} code, the density within the stream decreases as the material falls toward the primary star, such that $v\rho$ remains constant along the length of the stream. The coordinates of the origin and termination of the stream are specified within the code, and form the endpoints of the straight cylinder that represents the stream.  In this case, the stream extends from (35, 0)$R_{\odot}$ to (5, 5)$R_{\odot}$.  Similar results are expected from a finer geometric model of the stream that takes into account its curvature due to orbital motion, the full velocity field, and likely changing radius. The preferred parameters for the stream
are initial/final velocities of 100/500~km~s$^{-1}$, an initial
density of $6\times10^{-14}$~g~cm$^{-3}$, a temperature of 8000~K, a
radius of 4 $R_{\odot}$, and a turbulence velocity of 100~km~s$^{-1}$; with relatively large uncertainties of $\sim$15-25\%.

%%FIGURE 8 HERE

\begin{figure}
\figurenum{8}
\epsscale{1.2}
\plotone{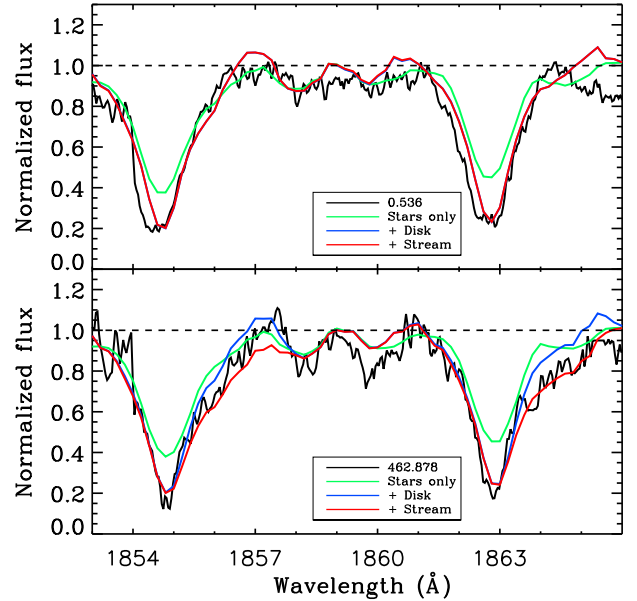}
\vspace{-0.5cm}
\caption{Models for \ion{Al}{3} compared to IUE data (black) at
${\phi}_{\rm 0}=0.536, 0.878$. The model with only stars (green) has
shallower central absorption than the data. Adding an accretion disk
(blue; see $\S$3.1) is a good match at ${\phi}_{\rm 0}=0.536$ but
overpredicts the red-wing emission at ${\phi}_{\rm 0}=0.878$; here,
better agreement is obtained when the gas stream (red; see $\S$3.2) is
also incorporated. The observed spectrum in the top frame is the average from phases ${\phi}_{\rm 0}=499.511$ and $591.560$.
}
\label{f8}
\end{figure}

The influence of the gas stream is apparent as absorption at
$0.75<{\phi}_{\rm 0}<1.0$, as discussed in $\S$2. For example,
inclusion of the stream has no appreciable impact on the modeled
\ion{Al}{3} line profile at low orbital phases, but Figure~8
illustrates that at ${\phi}_{\rm 0}=0.878$ the infalling stream,
viewed here against the primary, generates significant redshifted
absorption that improves the agreement between model and
observations.  We noticed that the blue side of the \ion{Al}{3} lines 
was variable near orbital phase 0.5; this is similar to the Si II 
variability observed by Peters (1994) which was interpreted as due to 
episodic mass loss. The final model provides a reasonable match to this 
variable \ion{Al}{3} line profile. 
The absorption at high orbital phases is less noticeable at H$\alpha$
since a mass-transfer stream will tend to absorb relatively more 
light at shorter wavelengths (e.g., Olson \& Bell 1998). However, at a temperature of
8000~K the stream also acts to provide additional H$\alpha$ emission
at other phases; for example, at ${\phi}_{\rm 0}=0.697$ the stream
generates slightly blueshifted emission that improves the agreement
between model and observations (Figure~6). Inclusion of the stream has
only a negligible effect for the H$\beta$ (Figure~7) and \ion{Si}{4} regions.

%%FIGURE 9 HERE

\begin{figure*}
\vspace{-0.7cm}
\figurenum{9}
\epsscale{0.9}
\plotone{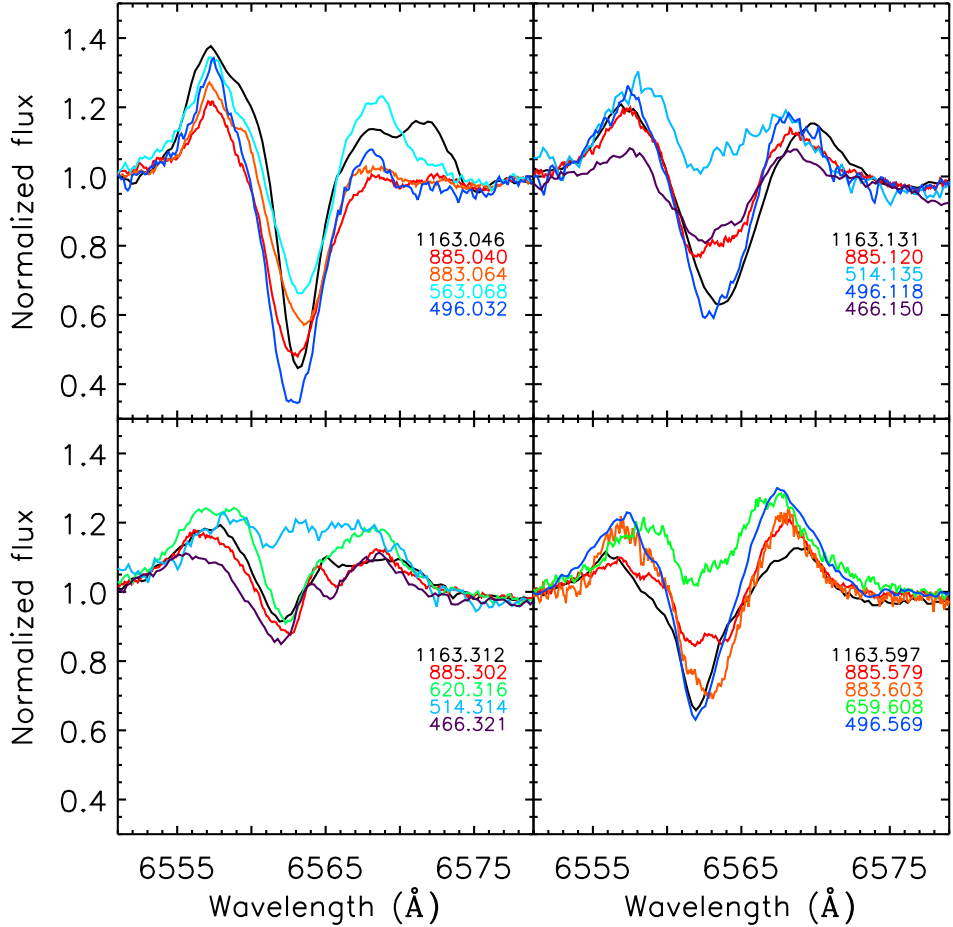}
\vspace{-0.7cm}
\caption{Long-term variability in the H$\alpha$ line profile.  The contrast over the 20-year period from 1991 to 2011, over 697 orbital cycles from epoch 466 to 1163, illustrates the extent of the line profile changes in AU Mon.  The variation over only two orbital cycles from epoch 883 to 885 demonstrates the rapidity with which the profile can change.
}
\label{f9}
\end{figure*}

The velocity and density adopted for the gas stream correspond to a
mass-transfer rate of $\sim$$2.4\times10^{-9} M_{\odot}$
yr$^{-1}$; this should be regarded as a lower limit since the loosely constrained temperature of the
stream is set near the peak emissivity of H$\alpha$.  Measurements of any changes in the period
over the last several decades would provide an independent estimate of
the mass-transfer rate in AU~Mon. The rate we derive here from {\it
Shellspec\/} modeling is broadly similar to mass-transfer estimates
for other long-period Algol-type binary systems, such as TT~Hya
($P$=6.95~d, $q$=0.27), for which Miller et al.~(2007) calculated a
rate of $\ge2\times10^{-10}M_{\odot}$~yr$^{-1}$, or the non-eclipsing
CX~Dra ($P$=6.70~d, $q$=0.23), for which Simon~(1998) calculated a
rate of $10^{-10}-10^{-8}M_{\odot}$~yr$^{-1}$.

%%%%%%%%%%%%%%%%%%%%%%%%%%%%%%%%%%%%%%%%%%%%%%%%%%%%%%%%%%%%%

\section{Spectral Variability}

\subsection{H$\alpha$ Variability}

The H$\alpha$ spectra described in this work reveal changes in
the emission line profile on both short (multi-week) and long
(multi-year) timescales over the 20-year period from 1991 to 2011 
and covering 697 orbital cycles (see Figure~9); this is consistent 
with previous observations (e.g., Plavec \& Polidan 1976; Sahade et al.~1997; 
Desmet et al.~2010). The HET data were taken
over $\sim$42 days ($\sim$3.8 orbital periods), with a 10-day gap
in the coverage. The 10 spectra obtained prior to this gap typically
show deeper central absorption than the 14 spectra obtained after the
gap. This is likely simply due to random short-term variability,
although it is alternatively possible that the decrease in central
absorption depth is partially related to the modest increase in the
long-period phase, which transitions from ${\phi}_{\rm L}\le22.83$ to
${\phi}_{\rm L}\ge22.86$ over the gap in coverage.   Figure~9 illustrates 
the rapidity with which the profile can change over only two orbital cycles 
(e.g., epoch 883 to 885).

We ran the grid of models on five subsets of H$\alpha$ spectra,
separated by long-period phase, to investigate potential changes in
the structure of the disk. The subsets were grouped as
$12.49\le{\phi}_{\rm L}\le12.51$ (Peters KPNO), $15.79\le{\phi}_{\rm
L}\le15.81$ (first Richards KPNO), $16.85\le{\phi}_{\rm L}\le16.86$
(second Richards KPNO), $22.80\le{\phi}_{\rm L}\le22.83$ (first HET),
and $22.86\le{\phi}_{\rm L}\le22.90$ (second HET). Although the
preferred model is better matched to the first set of HET spectra than
to the second, the change in central absorption does not
systematically alter the relative goodness-of-fit across the model
grid (i.e., generally the same subset of models remains favored). The
Peters KPNO spectra are also reasonably well-fit by the preferred
model and its similar cousins. In contrast, both sets of Richards KPNO
spectra have increased emission and are better fit by a stronger disk,
for example with a thickness of 6.4 $R_{\odot}$ and an outer radius of
25 $R_{\odot}$; it is additionally necessary to increase the stream
contribution through changing the density to
120$\times10^{-15}$~g~cm$^{-3}$, but
this is still not sufficient to provide a very good match to the
shallower central absorption. 

The excess low-velocity emission suggests the presence of additional material beyond that contained in the circularly symmetric disk and the stream (see tomography results in \S5); while the disk remains dominant, the residual emission seems more centralized. This behavior is reminiscent of CX Dra, for which Richards et al. (2000) concluded that the emission originates primarily between the stars as a result of the impact of the stream onto the outer edge of an extended disk.  It is also consistent with vertical outflows similar to the bipolar wind hypothesized in the double periodic variable V393 Sco (Mennickent et al. 2012). In any event, since the long-period phase is similar for
the Richards KPNO and the HET data, these results demonstrate that the long-period phase is not the sole driver of H$\alpha$ variability.

\subsection{Transient UV feature}

The model presented in $\S$3 including the stars, disk, stream, 
and a standard spot provides a good match to the majority of the
{\it IUE\/} observations. However, an additional secondary absorption 
feature is present in several spectra. This transient spectral
feature has the following characteristics: (i) the profile shape
appears to require some blueshifted absorption, rather than arising
exclusively from a spike of low-velocity emission; (ii) the feature is
only found in spectra obtained at $0.84<{\phi}_{\rm 0}<0.99$, although
most spectra at these phases do not show the feature; (iii) there is
no obvious correlation in the presence of this feature with the
long-period phase ${\phi}_{\rm L}$; and (iv) the feature is present in
5/13 cases in \ion{Si}{4} and in 2/13 cases also in \ion{Al}{3} and
\ion{Si}{2}.  (There is one exceptional case in which the line profile
is distorted at \ion{Al}{3} but not \ion{Si}{4}, at an intermediate
phase, but here the profile may be more consistent with excess
low-velocity emission.) This feature is unlikely to be related to any
potential calibration issues, since it has a consistent profile when 
it is present in both \ion{Si}{4} and \ion{Al}{3}. It is also unlikely 
to be of geocoronal or interstellar origin, since it is variable
and its presence is confined to a specific system phase.

The transient UV feature could in principle result from some
combination of blueshifted absorption and increased central
emission. For example, Bisikalo \& Kononov (2010) discuss a flare in
the CV SS~Cyg that generated strong H$\gamma$ emission while also
broadening the absorption. Another possibility is that the feature
results from a P-Cygni profile overlaying the stellar absorption,
although this would require material at larger radii (limiting the
temperature and hence the impact on \ion{Si}{4}). In AU~Mon, the
feature primarily modifies the blue side of the profile, and the
absorption then extends to shorter wavelengths than the disk plus
stream model predicts. We conjecture that the transient spectral
feature is associated with an occasional outflow arising in the
vicinity of the disk-stream interaction site, which is hot and of
variable temperature. It is possible that such an outflow could be
related to the HTAR described by Peters \& Polidan (1984, 1988) and
that both are consequences of the stream impacting the disk and
primary. In this context, the van Rensbergen et al.~(2011) result may 
be relevant: that Algols with hotter primaries (such as AU~Mon) may 
experience more mass loss than those with cooler primaries (such as TT~Hya).

%%FIGURE 10 HERE

\begin{figure}
\figurenum{10}
\epsscale{1.2}
\plotone{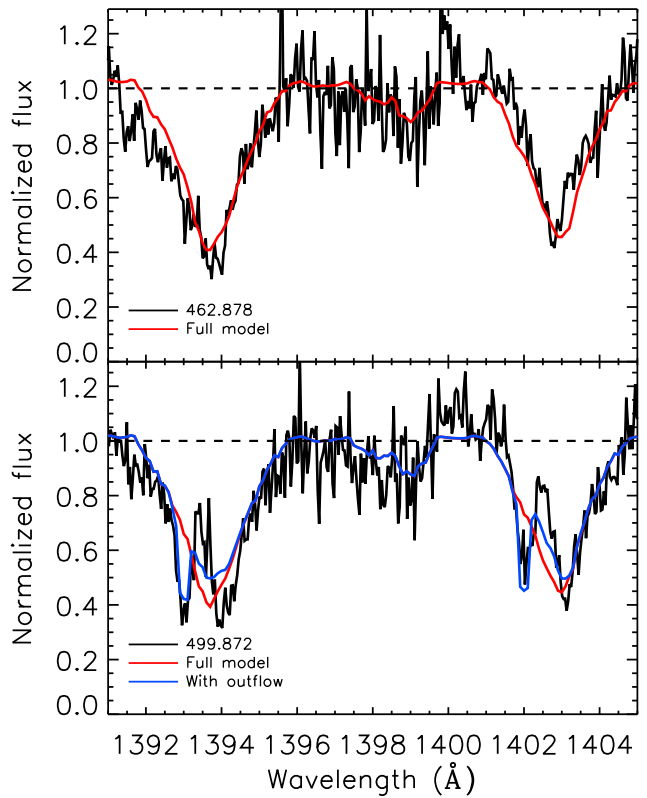}
\vspace{-0.9cm}
\caption{Models for \ion{Si}{4} compared to IUE data (black) near ${\phi}_{\rm 0} = 0.87$ for epoch 462 (top) and 499 (bottom).  A subset of the UV spectra, including at epoch 499 (but not 462), display a secondary blueshifted absorption feature that may indicate a transient outflow. (The \ion{Al}{3} profile at 499.872 shows a similar effect.) The standard model (stars,
accretion disk, gas stream, and spot/localized region) is shown in red, and a modified model with an added ``outflow'' spot component is shown in blue.  The extended model with the outflow component provides a better match to the observations.
}
\label{f10}
\end{figure}

We used {\it Shellspec\/} to verify that it is possible to produce such
an outflow feature (Figure~10) by using a second ``spot'' component in addition to the spot applied in the full model to represent the localized region between the stars; the extra spot is only used here to illustrate the transient UV feature. An ``outflow spot'' located at coordinates 
of (8, 6) $R_{\odot}$ with a velocity vector of
(120, 120)~km~s$^{-1}$ and with a spot radius/density/temperature of
$4R_{\odot}/30\times10^{-15}$~g~cm$^{-3}/12000$~K, significantly
improves the agreement between the modeled and observed UV spectra for
those cases in which this additional absorption feature is
present. The degree of improvement varies, as might be expected given
the variable nature of the feature itself. Although the outflow spot
parameters are based on the UV spectra, the addition of this
component would also improve the agreement between the modeled and observed
optical spectra in several cases. For example, both the H$\alpha$ and
H$\beta$ spectra at ${\phi}_{\rm 0}=883.236$ show features in the red
wing that are better matched when the transient outflow is
included; this effect may be partially due to the outflow spot component 
replacing disk emission within the overlapping region.

%%%%%%%%%%%%%%%%%%%%%%%%%%%%%%%%%%%%%%%%%%%%%%%%%%%%%%%%%%%%%
%%FIGURE 11 HERE

\begin{figure}
\figurenum{11}
\epsscale{1.15}
\plotone{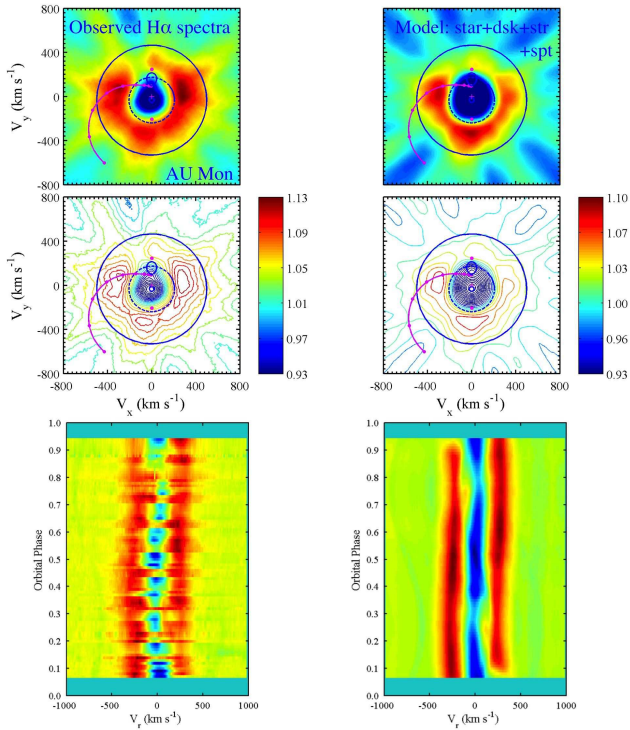}
\caption{Comparision between the observed spectra and the model based 
on the stars, disk, stream, and localized region (labelled ``spot'').  Left frames show the Doppler 
tomogram, contour map, and interpolated trailed spectrogram based on the 
observed spectra, while the right frames show the results based on the 
combined model for the stars, disk, stream, and spot.  The L2 and L3 points are represented by the large lavendar dots along the line of centers beyond the secondary and primary, respectively.  
The contour maps show that the model is fairly close to the observations.}
\label{f11}
\end{figure}

\section{Application of Doppler Tomography}

Doppler tomography provides an independent method of investigating the
accretion structure of AU~Mon, and of evaluating the accuracy of the
{\it Shellspec\/} modeling (described in $\S$3). 
The basic assumptions and constraints of Doppler tomography include (Marsh \& Horne 1988; Richards 2004): (1) the spectra are assumed to be broadened primarily by Doppler motions, (2) spectra with high wavelength resolution and good resolution in orbital phase (or projections) are needed to create a well-resolved image, (3) spectra dominated by emission lines are primarily used in the analysis, (4) the gas is assumed to be optically thin, and (5) the object
(velocity field+state quantities) does not change with time.  Several accretion structures have been identified in the 2D and 3D Doppler tomograms based on optical and ultraviolet spectra of Algol-type binaries (e.g., Richards 2004; Richards et al. 2012).  These include classical and transient accretion disks, the gas stream, emission from the chromosphere and other magnetic structures on the donor star (e.g., prominences and coronal mass ejections), shock regions, an accretion annulus, and an absorption zone (corresponding to a region of hotter gas). Moreover, tomography was used to demonstrate the quality of {\it Shellspec\/} models for the gas stream and disk in TT Hya, and to discover that the disk in TT Hya is asymmetric (Miller et al. 2007).  This procedure was extended for the study of AU Mon.

The observed spectra of AU Mon obtained over a range of phases were 
combined to construct a two-dimensional velocity-space map of the 
emission from the stars and accretion structures. Figure~11 shows the 
good match between the observed spectra and the model based on the 
stars, disk, stream, and a spot/localized region. The left frames show the
Doppler tomogram, contour map, and interpolated trailed spectrogram
based on the observed spectra, while the right frames show the results
based on the combined model for the stars, disk, stream, and spot. 
The observed tomogram shows that the strong H$\alpha$ emission arises 
primarily from the region in the tomogram that corresponds to the locus 
of a circular Keplerian accretion disk.  Since a non-Keplerian disk would 
have a range of lower velocities, the tomogram suggests that the disk is 
Keplerian.  The gap in the H$\alpha$ disk emission seen near $V_{\rm x}=0, V_{\rm y}>0$ is 
due to limited coverage at those phases (see Figure~1). The contour maps 
show that the model is fairly close to the observations.

Figure~12 shows the 2D Doppler tomograms obtained when the model spectra
for the various structures (stars, disk, stream, and spot) are
sequentially removed from the observed spectra.  This figure shows the
Cartesian representation of the binary (top left), the observed and
combined model tomograms (top right), the difference tomograms in
sequence (middle frames) for (1) observed - stars, (2)
obs-(stars+disk), (3) obs-(stars+disk+stream), and (4)
obs-(stars+disk+stream+spot).  The images for the latter three are
enhanced in the bottom frames. It is apparent that the H$\alpha$
emission in AU Mon is dominated by the accretion disk, then by the
stream, and spot (Localized Region, LR).  The relatively weak contribution
from the gas stream is apparent in the tomogram of the synthetic
spectrum constructed from the stars+disk+stream model, and in the
tomogram of the difference profile when the stars+disk model is
subtracted. The bulk of the stream emission occurs at velocities less
than $\sim$400~km~s$^{-1}$, corresponding to the relatively higher
density regions closer to the stream launch point at L1. This may reflect
the inner stream being subsumed within the disk as it approaches the
primary, which is modeled in {\it Shellspec\/} through prioritizing
the disk in overlapping regions.  

%%FIGURE 12 HERE

\begin{figure*}
\figurenum{12}
\epsscale{1.19}
\plotone{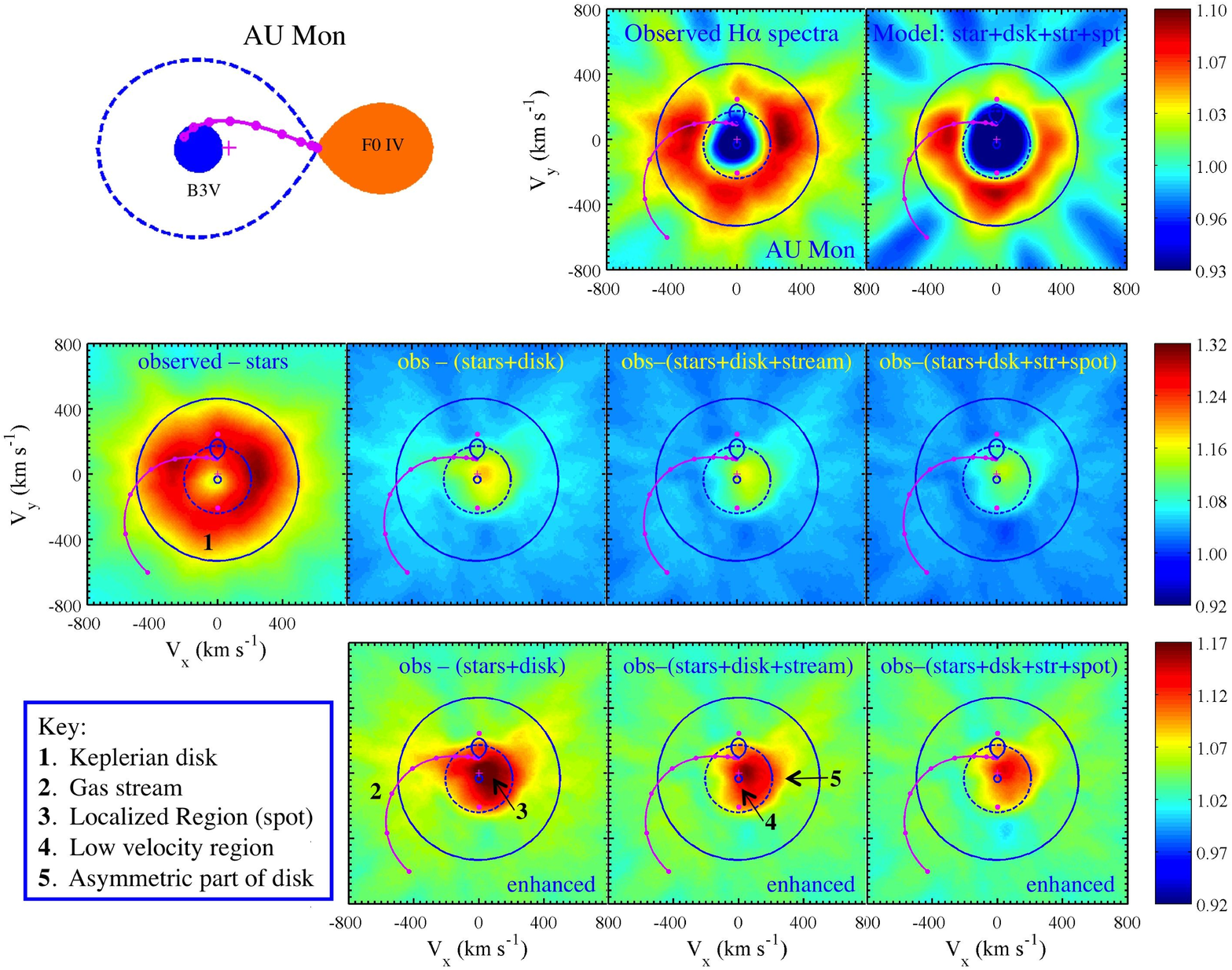}
%\vspace{0.1cm}
\caption{Two-dimensional Doppler tomograms obtained when the 
model fits for the various structures (stars, disk, stream, spot/localized region) are 
sequentially removed from observed spectra.  This figure shows the 
Cartesian representation of the binary (top left), the observed and 
combined model tomograms (top right), the difference tomograms in 
sequence (middle frames) for (1) observed - stars, (2) obs-(stars+disk), 
(3) obs-(stars+disk+stream), and (4) obs-(stars+disk+stream+spot).  The 
images for the latter three are enhanced in the bottom frames.}
\vspace{0.3cm}
\label{f12}
\end{figure*}

The images show that we have successfully modeled the dominant part of
the disk, the gas stream, and the brightest part of the LR spot feature.  
However, the final image shows that there are still two parts
of the image that remain: a low velocity region near the LR and also
an asymmetric portion of the disk. The location of this residual
emission in velocity space is broadly similar to that observed in
TT~Hya, which Miller et al.~(2007) ascribed to an asymmetric disk, as
supported by hydrodynamic simulations by Richards \& Ratliff~(1998);
it seems plausible that the disk in AU~Mon is also asymmetric. 
Moreover, compared to TT~Hya, the residual emission in AU~Mon is more 
pronounced and has clearly sub-Keplerian velocities.  

A possible explanation for these lower velocities is that the stream in AU Mon makes a nearly tangential collision with the primary (see Figure~1), reducing the circular speed of the continuing flow, which could then contribute additional low-velocity emission, particularly along the line of centers where the density rises due to conservation of particle flux as it collides with the incoming gas stream.  In the case of TT~Hya, the system geometry precludes the stream from a direct collision with the primary.   Given the early spectral type of the primary in AU Mon, it is also possible that the residual emission seen in the tomogram could be explained by a vertical wind or jet as found in $\beta$ Lyr (Ak et al. 2007) and V393 Sco (Mennickent et al. 2012).  The 2D tomograms displayed in Figures~11 and 12 (top frames) suggest that the disk in AU Mon extends almost to the Roche surface of the primary (since the inner disk velocity is similar to that of the Roche surface), so any subsequent gas flow from the L1 point would be intercepted by this extended disk structure and perhaps create a jet-like feature.  A similar extended disk structure was also found in CX Dra (Richards et al. 2000).  These features might be confirmed later through the application of 3D tomography.  In the meantime, it is encouraging that a modified {\it Shellspec\/} model with an added outflow spot component beyond the full synthetic spectrum based on the stars+disk+stream+spot/LR can provide a better match to the observations than when the outflow region is excluded; as illustrated in the case of the \ion{Si}{4} line (see Figure~10). 

The tomograms do not show any obvious evidence of mass loss or depletion in the emission 
associated with the L2 and L3 points (see large lavendar dots along the line of centers, beyond the stars in Figure~12).  However, if mass loss has occurred or is in progress, then the tomograms confirm that there is a greater likelihood that the source should be located near the L1 point, as suggested by Barr{\'{\i}}a \& Mennickent~(2011).

Although the H$\alpha$ central absorption changes in depth throughout the long-period cycle
(Figure~3), the residual emission revealed in the final frame of Figure~12 is present for both
$0.35<{\phi}_{\rm L}<0.85$ and $0.85<{\phi}_{\rm L}<1.35$, confirming
that this effect is tied to the relative geometry of the stars, disk,
and stream, rather than arising from long-period variability such as
due to cyclic changes in the mass transfer rate.

%%%%%%%%%%%%%%%%%%%%%%%%%%%%%%%%%%%%%%%%%%%%%%%%%%%%%%%%%%%%%
%\newpage
\section{Summary}

The main results from our observations and modeling of accretion
structures in AU~Mon are the following:

1. We confirm that the H$\alpha$ and UV spectral features display
significant variability, some of which is linked to the long-period
phase. For example, spectra taken within $0.35<{\phi}_{\rm L}<0.85$
show deeper central absorption than those taken within ${\phi}_{\rm
  L}>0.85$ or ${\phi}_{\rm L}<0.35$. We find that these trends are
broadly consistent with the scenario of variable mass transfer
proposed by Peters~(1994).

2. We use {\it Shellspec\/} to self-consistently model the accretion
disk and gas stream. Preferred parameters for the disk and stream are
determined through comparison to observed spectra; for the accretion
disk, they include a thickness of 5.2 $R_{\odot}$, an outer radius of
23 $R_{\odot}$, a maximum temperature of $\simeq$14000~K, and a density
of $10^{-13}$~g~cm$^{-3}$, and for the stream they include
initial/final velocities of 100/500~km~s$^{-1}$, an initial density of
$6\times10^{-14}$~g~cm$^{-3}$, a temperature of 8000~K, and a radius
of 4 $R_{\odot}$. A lower limit for the mass transfer rate was found to
be $\sim$$2.4\times10^{-9} M_{\odot}$ yr$^{-1}$.

3. We generate Doppler tomograms from the observed and difference
spectra to investigate the two-dimensional accretion structure in 
velocity space. Sequential removal of the stellar/disk/stream/spot
components confirms the accuracy of the {\it Shellspec\/} modeling,
and additionally identifies unmodeled aspects of the system.  These include
an asymmetric component of an elliptical accretion disk, and material
moving at sub-Keplerian velocities that provides excess H$\alpha$
emission, plausibly associated with the continuation of the mass
transfer stream beyond the splash site or perhaps an source of outflow 
or mass loss from the system.

\acknowledgments
We thank the referee for helpful comments on the manuscript. This research was partially supported by NSF grant AST-0908440 (MTR)
and NASA ADP grants NNG04GC48G and NNX12AE44G (GJP and MTR). The
Hobby-Eberly Telescope (HET) is a joint project of the University of
Texas at Austin, the Pennsylvania State University, Stanford
University, Ludwig-Maximilians-Universit{\"a}t M{\"u}nchen, and
Georg-August-Universit{\"a}t G{\"o}ttingen. The HET is named in honor
of its principal benefactors, William P. Hobby and Robert
E. Eberly. The Image Reduction and Analysis Facility (IRAF: {http://iraf.noao.edu/}) was used in this work, and the MATLAB software package was used to make the tomography images.   The latest version of the {\it Shellspec\/} code can be obtained at {http://www.astro.sk/$\sim$budaj/shellspec.html}.

\clearpage

\end{document}